\documentclass[preprint,authoryear]{elsarticle}

\usepackage{placeins}
\usepackage{subcaption} 
\usepackage{graphics}
\usepackage{wrapfig}
\usepackage{hyperref}
\usepackage{multicol}
\usepackage{amsmath,amsfonts,amssymb}
\usepackage{graphicx}
\usepackage{setspace}
\usepackage{tocloft}
\usepackage{scrextend}
\usepackage{hhline}
\usepackage{tabularx}
\usepackage[utf8]{inputenc}
\usepackage{multirow}
\usepackage{mathrsfs}
\usepackage{adjustbox}
\usepackage{bbm}
\usepackage{tablefootnote}
\usepackage{enumitem}
\usepackage{amsthm}

\newtheorem{example}{Example}[section]

\usepackage{lineno}

\usepackage{xcolor}

\usepackage{booktabs}
\usepackage[ruled,vlined]{algorithm2e}
\usepackage{algorithmic,float}
\def\arrvline{\hfil\kern\arraycolsep\vline\kern-\arraycolsep\hfilneg}
\usepackage{scalerel,stackengine}
\graphicspath{ {./Bilder/} }
\stackMath
\newcommand\reallywidehat[1]{%
\savestack{\tmpbox}{\stretchto{%
  \scaleto{%
    \scalerel*[\widthof{\ensuremath{#1}}]{\kern-.6pt\bigwedge\kern-.6pt}%
    {\rule[-\textheight/2]{1ex}{\textheight}}
  }{\textheight}%
}{0.5ex}}%
\stackon[1pt]{#1}{\tmpbox}%
}

\SetCommentSty{mycommfont}

\SetKwInput{KwInput}{Input}                
\SetKwInput{KwOutput}{Output}              

\usepackage{multirow}

\title{From point forecasts to multivariate probabilistic forecasts: The Schaake shuffle for day-ahead electricity price forecasting }

\author[1]{Oliver Grothe}
\author[1]{Fabian Kächele \corref{cor1}}
\author[2]{Fabian Krüger}
\ead{fabian.kaechele@kit.edu}
\address[1]{Karlsruhe Institute of Technology (KIT), Institute of Operations Research, Analytics and Statistics, Kaiserstr. 12, 76131 Karlsruhe, Germany.}
\address[2]{Karlsruhe Institute of Technology (KIT), Institute of Economics, Applied Econometrics, Kaiserstr. 12, 76131 Karlsruhe, Germany.}
\cortext[cor1]{Corresponding author\\
\textit{Email address:} \href{mailto:fabian.kaechele@kit.edu}{fabian.kaechele@kit.edu}}

\begin{document}
\journal{Energy Economics}

\begin{abstract}
Modeling price risks is crucial for economic decision making in energy markets. Besides the risk of a single price, the dependence structure of multiple prices is often relevant. We therefore propose a generic and easy-to-implement method for creating multivariate probabilistic forecasts based on univariate point forecasts of day-ahead electricity prices. While each univariate point forecast refers to one of the day's 24 hours, the multivariate forecast distribution models dependencies across hours. The proposed method is based on simple copula techniques and an optional time series component. We illustrate the method for five benchmark data sets recently provided by \cite{lago2020forecasting}. Furthermore, we demonstrate an example for constructing realistic prediction intervals for the weighted sum of consecutive electricity prices, as, e.g., needed for pricing individual load profiles.
\end{abstract}

\flushbottom
\maketitle
\thispagestyle{empty}
\section{Introduction}
Day-ahead electricity price forecasting is a critical element in the decision-making of energy companies. Accordingly, an active applied research literature is concerned with developing and comparing price forecasting methods. The majority of this literature addresses point forecasts, for which a wide range of methods has been proposed. These efforts, which have recently been reviewed by \cite{9218967} and \cite{lago2020forecasting}, have given rise to a rich toolkit that includes time series models, regularized regression techniques, deep learning models, as well as strategies for parameter estimation, variable selection, hyperparameter selection and forecast combination.\\

While univariate point forecasts are an important first step, they are limited in two respects. First, they do not address forecast uncertainty, which is an important concern for economic decision making. The studies surveyed by \cite{NOWOTARSKI20181548} and \cite{ZIEL2018251} address this aspect by considering probabilistic forecasting in energy markets. While growing, the corresponding literature and range of approaches is still more limited than for point forecasting. Second, univariate forecasts are concerned with a single measurement unit (such as a given time period or a given location), whereas economic decisions often require joint forecasts for several time periods or locations. In applications, the two aspects of forecast uncertainty and multivariate dependence may well interact. For example, managing daily price risks requires information on both the uncertainty in hourly prices and information on dependencies of prices across hours. Consider the sum of the prices from two consecutive hours. If the price of the first hour is higher than predicted, the price of the second predicted hour is likely to be higher as well. Assuming independence of these prices would underestimate the uncertainty of the sum, leading to suboptimal economic decisions. Dependence modeling thus plays a vital role in uncertainty estimation.\\ 

This paper considers a generic and easy-to-implement method for constructing multivariate probabilistic forecasts based on univariate point forecasts. The method uses past forecast errors to learn about forecast uncertainty and dependencies across measurement units. If necessary, we first fit simple time series models to the univariate point forecast errors, thus accounting for possible predictability in the errors' conditional mean and variance. For dependence, we apply straightforward empirical copula methods. Being based on forecast errors, our approach leverages the rich literature on univariate point forecasting (thus avoiding to re-invent the wheel) while at the same time addressing forecast uncertainty and multivariate dependence. 
It is inspired by the literature on post-processing multivariate ensemble forecasts in meteorology \citep{TheSchaakeShuffle,schefzik2013,VannitsemEtAl2021}. Here, the point forecast errors of numerical weather prediction models are used to learn about forecast uncertainty and (possibly) about multivariate dependencies across variables, locations or time points. Related post-processing approaches have been considered in various other areas of application, such as macroeconomics \citep{KruegerNolte2016,ClarkEtAl2020}. \\

Post-processing seems particularly promising for forecasting energy prices or demand, given the rich literature on point forecasting and the wide availability of data for learning structures in past point forecasting errors. Nevertheless, the literature on post-processing of energy forecasts is relatively sparse. In a univariate context, studies such as \cite{MARCJASZ2020466} and \cite{KATH2021777} discuss approaches to estimate forecast uncertainty from past point forecast errors. \cite{PhippsEtAl2020a,PhippsEtAl2020b} and \cite{LudwigEtAl2020} consider post-processing univariate or multivariate weather forecasts, in a situation where these are used as an input to energy forecasting models. \cite{MuniainZiel2020} deal with bivariate probabilististic price forecasting, using a residual based approach that is conceptually similar to post-processing. \cite{Janke_2020} apply an implicit generative model to generate a multivariate forecast distribution for energy prices from an ensemble of univariate point forecasts. Furthermore, \cite{2019Chai} employ a Gaussian copula to generate scenarios on the basis of an ensemble from extreme learning machines. Compared to the latter two studies, our use of empirical copula methods appears considerably simpler to implement and more easily comparable to the large meteorological literature. Other uses of copula methods to model multivariate dependence in energy contexts include \cite{2019Toubeau}, \cite{PINSON201212}, \cite{MANNER2016255} and \cite{PIRCALABU2017283}.\\

We next present a more specific summary of our approach, full details of which are given in Section \ref{sec:methodologies}. We consider a vector of $24$ day-ahead point forecasts, each of which refers to one hour $h$ of the following day $t$ (see \cite{lago2020forecasting} for a description of day-ahead markets). From now on, we stick to this hourly example to simplify the presentation, but quarter-hourly or minute-by-minute forecasts could be handled analogously. The prediction is subject to an error defined by $  \epsilon_{t,h}=y_{t,h} -\hat{y}_{t,h},
\label{epsilon rechnen}$ where $y_{t,h}$ denotes the true price and $\hat{y}_{t,h}$ the prediction for each day $t$ and hour $h=1,\dots,24$. Conditional on the information set $\mathcal{F}_{t-1}$, available at day $t-1$, $\epsilon_{t,h}$ is a random variable with unknown, conditional distribution function that we model as $G_{t,h}(x) := F_h( \frac{x - \mu_{t,h}}{\sigma_{t,h}}).$ Here, $F_h$ reflects an hour specific shape of the distribution while $\sigma_{t,h}$ and $\mu_{t,h}$ are scale and location parameters.\\

In what we call the \textit{Error Learning Phase}, the error distributions as well as the scale and location parameters are estimated. We allow a potentially time-varying conditional mean $\mu_{t,h} = \mathbb{E}[\epsilon_{t,h}|\mathcal{F}_{t-1}]$ and variance $\sigma_{t,h}^2 = \mathbb{V}[\epsilon_{t,h}|\mathcal{F}_{t-1}].$
The nonzero conditional mean $\mu_{t,h}$ is motivated by possible misspecification of the point forecasting model, leading to predictable forecast errors. Time variation in the conditional variance $\sigma^2_{t,h}$ accommodates for differences in price uncertainty over time. If predictable structure in the data is apparent, $\mu_{t,h}$ and $\sigma^2_{t,h}$ can be tracked by time series models, e.g., by estimating the models' parameters in rolling windows of past forecast errors, separately for each hour. Alternatively, with no predictable structure apparent, we simply set $G_{t,h}(x):=F_h(x)$, so that the error distribution remains constant over time. We call the latter the `raw'-error approach. In the \textit{Dependence Learning Phase}, the procedure aims to learn the dependence of the point forecast errors. Therefore, we estimate the joint distribution of the 24-dimensional vector of day-ahead forecasting errors $\epsilon_{t,h}$. We disentangle univariate error distributions and their dependence by inferring the copula from the corresponding uniformly standardized values, i.e., $\hat u_{t,h}:=\hat G_{t,h}(\epsilon_{t,h}),$ where $\hat G_{t,h}$ has been obtained from the \textit{Error Learning Phase}. Finally, the learned error distribution and dependence structure are combined in the \textit{Forecasting Phase} to construct a forecast distribution consisting of $m$ simulated time paths of day-ahead electricity prices, with each path consisting of $24$ hourly prices. 
This simulated distribution encodes both forecast uncertainty for each individual hour, and the temporal dependence across the day's $24$ hours. Among others, it can be used to assess the likelihood of price-related events (such as the event of a price spike), and to measure the uncertainty of summary random variables (such as the sum of hourly prices). We consider an example for the price of a Standard Load Profile, reflecting a typical daily pattern of energy consumption, and show that the proposed method is able to capture the dependence structure of hourly prices, leading to a realistic assessment of price risks. Furthermore, we highlight the importance of the dependence structure in a multivariate probabilistic forecast by comparing our results to their na\"{i}ve counterpart that assumes temporal independence of the prices.\\

The remainder of the paper is structured as follows. Section \ref{sec:data} presents the data sets of \cite{lago2020forecasting} and day-ahead load data that we consider in addition. In Section \ref{sec:methodologies} we explain our proposed method in detail and give a toy example. Section \ref{sec:results} provides a case study for multivariate probabilistic electricity price forecasting in five markets and an example for the price of a Standard Load Profile in Germany. Further, we apply our method on a highly seasonal load forecasting time series to show the benefit of the time series component. Last, Section \ref{sec:conclusion} summarizes our work and points towards future research.

\section{Data and Point Forecasts}
\label{sec:data}

\subsection{Data Sets}
\begin{figure}[ht]
    \centering
\includegraphics[width=1\linewidth]{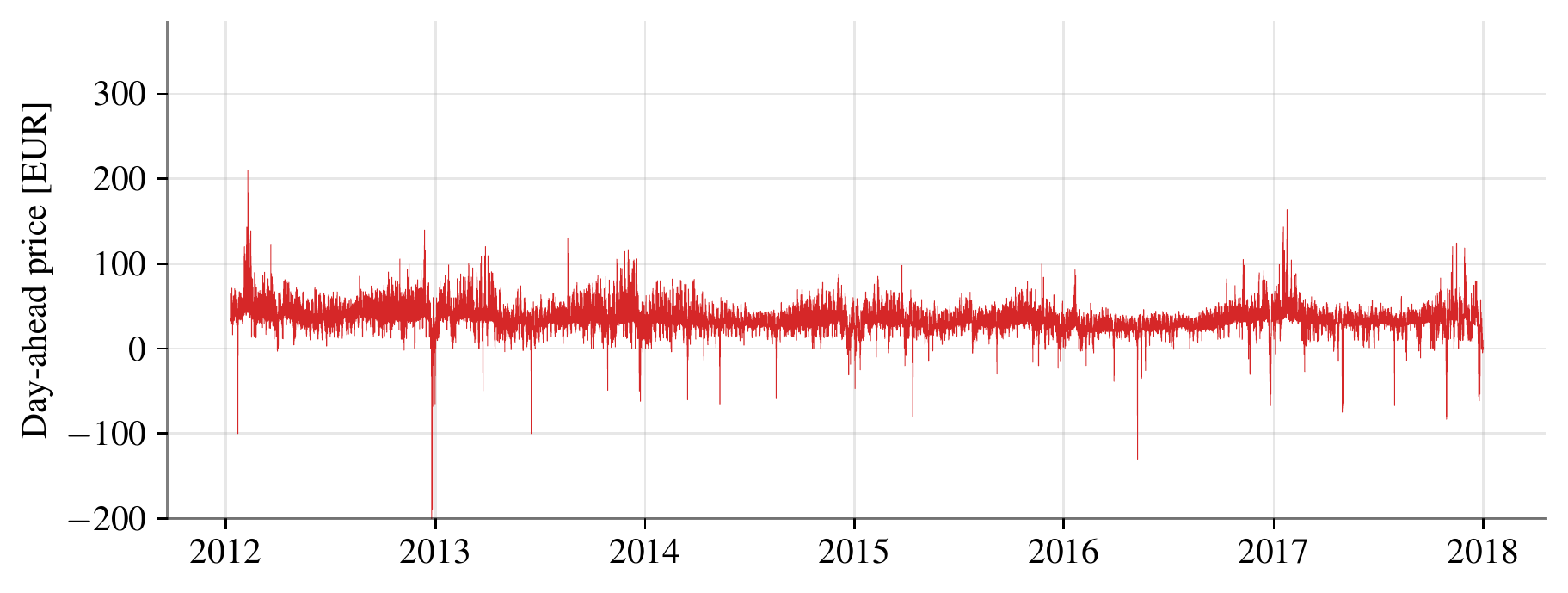}
    \caption{Day-ahead electricity prices in EUR/MW from 4 January 2012 to 31 December 2017 in the EPEX-DE dataset  \citep{lago2020forecasting}. }
    \label{DE_interval}
\end{figure}

The method we consider is based on a data set of past point forecasts and corresponding realizations. In our illustrations, we utilize point forecasts of energy prices and the corresponding realized prices. As discussed in the introduction, a wide range of point forecasting models and data sources are available for this purpose. Here we consider the five benchmark data sets recently made available by \cite{lago2020forecasting}. The data covers five day-ahead electricity markets, each spanning a history of six years. In addition to observations of day-ahead prices, two time series of influential exogenous variables are included, which differ for each market. To illustrate the data, Figure \ref{DE_interval} displays the realized day-ahead electricity prices in EUR per MW from 4 January 2012 to 31 December 2017 in the data set for the German electricity market (EPEX-DE).
\FloatBarrier We generate point forecasts for the five data sets from the implementation of the LEAR model by \cite{lago2020forecasting} with a two year calibration window.\footnote{Forecast averaging across calibration windows of different lengths could possibly yield improved results (\citealt{8458131,en11092364}); here we consider the simpler choice of a single calibration window in order to retain focus on the proposed method.} The LEAR model is a parameter-rich ARX model (that is, an autoregressive model with a large number of `exogenous' regressors or features) estimated using LASSO for feature selection. In total, 247 features are considered, including day-ahead prices, day-ahead forecasts of the two exogenous variables, and historical day-ahead forecasts of the exogenous variables, all stemming from previous days and weeks. Additionally, a dummy variable for the day of the week is included. Finally, all variables are preprocessed with the arc hyperbolic sine (asinh) variance stabilizing transformation. Details on the data and the LEAR model can be found in \citet[Section 4.2]{lago2020forecasting}.\\
\begin{wrapfigure}[14]{rt}{0.5\textwidth}
  \begin{center}

    \vspace*{-0.8cm} \includegraphics[width=0.5\textwidth]{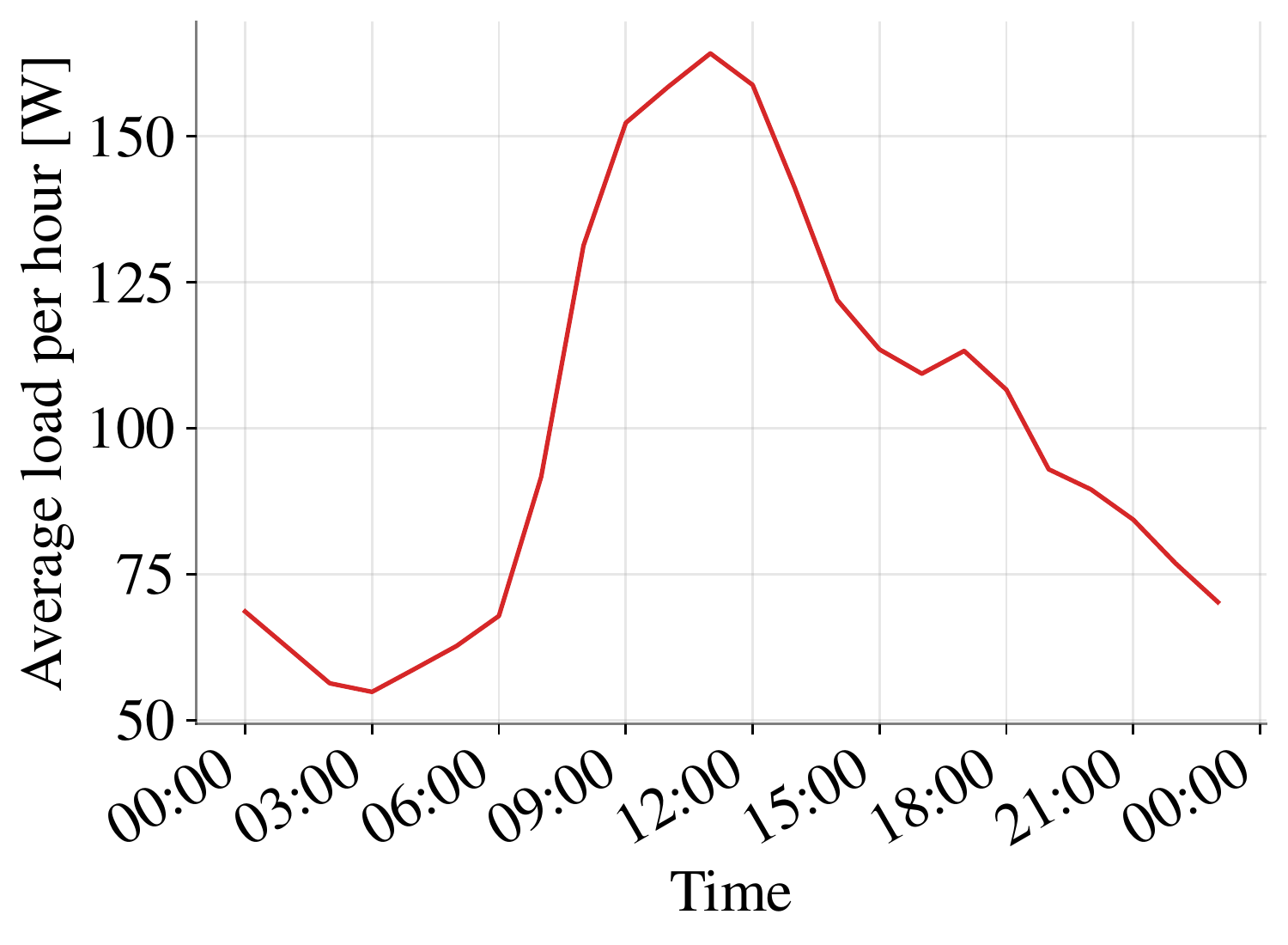}
  \end{center}
  
  \vspace*{-0.3cm}\caption{Load profile (SLP G0), averaged over all seasons, weekdays, and hours in [kW] for a normalized consumption of 1.000kWh/year.}
  \label{SLP}
\end{wrapfigure}In our forecast evaluation analysis, we further consider a \textit{Standard Load Profile (SLP)} that provides a practically relevant example of intra-day seasonality in energy demand. We use this profile to construct a single daily energy price from $24$ hourly prices. This setup allows us to assess whether our multivariate probabilistic predictions of hourly prices enable a realistic estimation of the uncertainty in daily prices. In practice, German energy suppliers use SLPs to model the consumption patterns of electricity customers without registered power metering. The SLPs replace the non-existent load profile of these customers with a forecast of electricity consumption during every quarter-hour. We utilize SLP G0 (see Figure \ref{SLP}), which represents the average of all industrial SLPs in Germany and is provided by the German Association of Energy and Water Industries \citep{BDEW}.\\

\begin{figure}[ht]
    \centering
\includegraphics[width=1\linewidth]{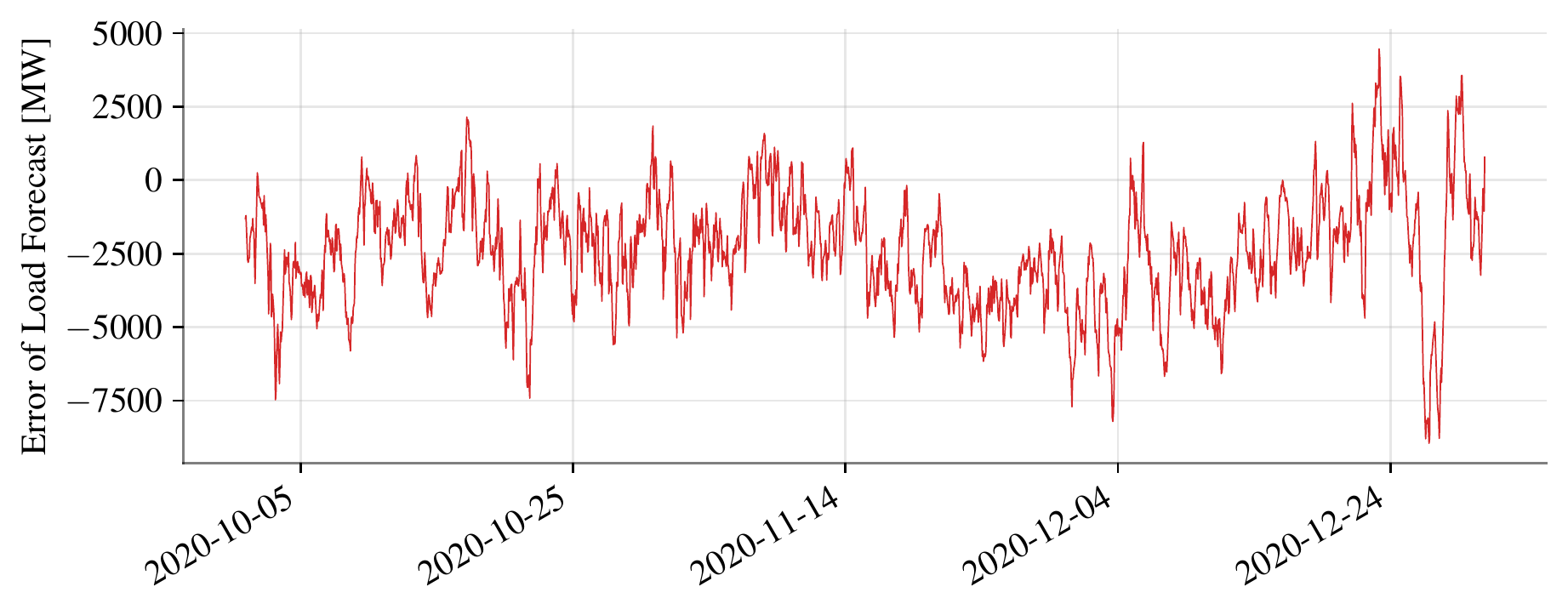}
    \caption{Load forecast errors for Germany from 1 October 2020 to 31 December 2020.  }
    \label{entsoe_figure}
\end{figure}

We also demonstrate the application of our method for the related task of load forecasting. To this end, we use day-ahead load data from the entso-e transparency platform (\url{https://transparency.entsoe.eu/}) for the German market from 2016 to 2020. Load forecasting is essential for planning and operation of energy suppliers, system operators and other market participants \citep{Weron}. Figure \ref{entsoe_figure} shows the time series of errors that results from a `black box' forecast provided on the platform. Upon closer inspection, the error series displays strong seasonal patterns and therefore calls for appropriate time series modeling techniques for postprocessing. 


\subsection{Sample Choices for Forecasting Scheme}
\label{rollwin}
Training of our model is based on two phases. First, the univariate error distribution is learned (\textit{Error Learning Phase}). Second, the joint behavior of realized and standardized errors is modeled by their copula (\textit{Dependence Learning Phase}). Both phases require calibration windows for learning and we implement the proposed method using backward-looking rolling windows. The windows should be long enough to provide stable estimates but also short enough to adapt quickly to changing market developments. Here, we use about one year (364 days) of raw errors $\epsilon_{t,h}$ for the Error Learning Phase. Then, the raw errors are standardized with the help of an appropriate time series model, and the most recent 90 standardized errors are used to infer the distribution of standardized errors $F_{h}$. Alternatively, when the raw errors do not show temporal structures, the $24$ distributions $F_{h}, h = 1, \ldots, 24$ are inferred directly from the most recent 90 raw errors without filtering them through a time series model. We call the latter `raw-error' approach. Finally, the Dependence Learning Phase estimates multivariate dependence by computing the copula between the 24 distributions $F_{h}$ based on a windows of the last 90 filtered or raw errors, respectively. While the trade-off between too short or too long window sizes needs to be resolved individually for each use case, our experiments indicate that the lengths proposed above are a good baseline setting for hourly prices.\\

After the next day's distribution of observations is predicted, the windows are rolled over by one day and the estimates of the error distributions and dependency structures are updated. The initial rolling windows used for each data set are shown in Table \ref{timeTable} along with the overall resulting evaluation periods.


 \begin{table}[ht]
 \footnotesize
     \centering
     \begin{tabular}{lccc}

\hline
         Dataset    & Error Learning   & Dependence Learning & Evaluation Period   \\
              \hline
         EPEX-DE    & 05.01.2015-03.01.2016         & 06.10.2015-03.01.2016 &04.01.2016-31.12.2017\\
         PJM        & 29.12.2015-26.12.2016         & 28.09.2016-26.12.2016 & 27.12.2016-24.12.2018 \\
         EPEX-BE    & 05.01.2014-03.01.2015         & 06.10.2014-03.01.2015 &04.01.2015-31.12.2016 \\
         EPEX-FR    & 05.01.2014-03.01.2015        & 06.10.2014-03.01.2015 &04.01.2015-31.12.2016 \\
         NORD POLE  & 29.12.2015-26.12.2016         & 28.09.2016-26.12.2016 &27.12.2016-24.12.2018\\
          \hline
     \end{tabular}
     \caption{Sample choices in the empirical analysis. We show the initial backward looking windows for the learning phases which are then rolled over until the end of the evaluation periods (rightmost column).}
     \label{timeTable}
 \end{table}
\section{Methods}
\label{sec:methodologies}

The method we propose is inspired by \cite{TheSchaakeShuffle} and \cite{schefzik2013} who consider multivariate dependencies in weather forecasting models, e.g., across locations or across different weather variables. Intuitively, we learn the univariate forecast distributions of 24 hours of the day (\textit{Error Learning Phase}) and their dependence structure (\textit{Dependence Learning Phase}) in history and use this information to construct multivariate forecasts in the future (\textit{Forecasting Phase}). We next detail the three steps of the method. An algorithmic description is available in the appendix (Algorithms \ref{algo1} and \ref{algo2}).

\subsection{Error Learning Phase}
In the \textit{Error Learning Phase}, we estimate the conditional distributions $G_{t,h}(x):=F_h( \frac{x - \mu_{t,h}}{\sigma_{t,h}})$ of the forecast errors ${\epsilon}_{t,h}$, where the indices represent the day ($t$) and hour ($h$). The conditional mean and variance of the error, $\mu_{t,h}$ and $\sigma_{t,h}^2$, can be estimated via a time series model. This is recommended if, e.g., the point prediction has high autocorrelation, seasonal effects or heterogeneous variance patterns. The estimates $\hat \mu_{t,h}$ and $\hat \sigma_{t,h}$ then result from the forecast of the time series model. The required degree of complexity of the model partly depends on the point forecasting model that generated this sequence of forecast errors. In an idealized setup, one would not require a model for the expected forecast error $\mu_{t,h},$ as the latter is equal to zero if the forecast model is correctly specified \citep[see e.g.][Section 2.3]{PesaranWeale2006}. However, due to complex seasonality and persistence patterns, this goal is hard to achieve for practical energy price forecasts \citep[see e.g.][]{MACIEJOWSKA2021105273}. Time variation in $\sigma_{t,h}^2$ reflects different market phases and seems unrelated to the accuracy of the point forecasting model.\footnote{To see this point, consider a stylized example where the true conditional distribution of $y_{t,h}$ given $\mathcal{F}_{t-1}$ has mean 5 and variance $1 + \mathbf{1}(t > 100),$ i.e., the conditional variance equals one for time periods $t \le 100$ and $2$ for $t > 100$. In this example, the optimal mean forecast of $y_{t,h}|\mathcal{F}_{t-1}$ is given by $5$, and the conditional variance of $\varepsilon_{t,h} = y_{t,h} - 5$ is heteroskedastic even though the mean forecast is optimal.} If no time series model is applied, we set $G_{t,h}(x)=F_h(x)$, i.e., we assume that the distribution of $y_{t,h}|\mathcal{F}_{t-1}$ does not change over time.\\  

To complete the model specification, we must estimate or fix the distribution $F_h$ of the standardized residuals $\hat z_{t,h} = \frac{\epsilon_{t,h} -\hat{\mu}_{t,h}}{\hat{\sigma}_{t,h}}$. We consider two versions: A parametric variant, where we suppose that $\hat z_{t,h}$ follows a standard normal distribution, and a non-parametric variant, where we estimate $F_h$ via the empirical distribution function. We hence set 
\begin{align}
    \hat{F}_h(z)=\frac{1}{n}\sum_{t=1}^{n}\mathbf{1}(\hat{z}_{t,h}\leq z),
\end{align} 
where $z \in \mathbb{R}$, and $n$ is the length of the error learning phase. 

\subsection{Dependence Learning Phase}
\label{Depmethod}

The \textit{Dependence Learning Phase} aims to learn the joint distribution of all $\epsilon_{t,h}$. Since we have already estimated the marginal distributions $G_{t,h}(x)$ of each hour $h$, we next estimate their dependence structure using copula techniques. To motivate our use of copulas, note that we need the joint distribution of the 24 day ahead errors $\epsilon_{t,h}$ of the point forecast model for $h=1,\dots,24$. 
This joint distribution is represented by the multivariate distribution function
\begin{align}
   G_\mathbf{\epsilon_t}(\theta)=P(\epsilon_{t,1} \le \theta_1,\dots,\epsilon_{t,24} \le \theta_{24})
   \label{Sklar}
\end{align}
with $\mathbf{\epsilon_t}=(\epsilon_{t,1},\dots,\epsilon_{t,24})$, where $\theta = (\theta_1,\ldots,\theta_{24}) \in \mathbb{R}^{24}$ is a vector of threshold values. For example, for a vector of zero thresholds, $\theta_0 = (0, \ldots, 0)$, $G_\mathbf{\epsilon_t}(\theta_0)$ yields the probability that all $24$ forecast errors are negative (or zero). In general, the distribution function $G_\mathbf{\epsilon_t}$ is a complicated object that depends on both the marginal distribution of each element $\epsilon_{t,h}$ as well as the dependence structure across the $24$ elements. However, copula methods allow us to specify the 24 elements' marginal distributions and their dependence separately. \cite{Nelson1998}, \cite{Joe}, and \cite{Priciples} provide textbook introductions to copulas. By Sklar's theorem, there exists a copula function $C: [0,1]^{24} \rightarrow [0,1]$ such that 
\begin{align}
  G_\mathbf{\epsilon_t}(\theta) = C(G_{t,1}(\theta_1), \ldots, G_{t,24}(\theta_{24})),
\end{align}
where the inputs are the elements' marginal CDFs, $G_{t,h}(\theta_h) = P(\epsilon_{t,h} \le \theta_h) \in [0,1]$ \citep{Skla59}. That is, the copula $C$ constructs a multivariate CDF from 24 marginal CDFs. A simple example of a copula is given in the case of independence, where $C(G_{t,1}(\theta_1), \ldots, G_{t,24}(\theta_{24})) = \prod_{h=1}^{24} G_{t,h}(\theta_h),$ i.e., the copula function simply multiplies all of its arguments. In practice, independence of the hourly prices is clearly unrealistic and more flexible models of dependence are required. We can achieve this by choosing and estimating an appropriate copula function $C$. Both parametric and non-parametric methods are available for this purpose. In the parametric case, we require additional parameters that determine the function $C$. As a prominent example, the Gaussian copula sets
\begin{align} 
C(G_{t,1}(\theta_1), \ldots, G_{t,24}(\theta_{24})) = \mathbf{\Phi}_{\Sigma}(\Phi^{-1}(G_{t,1}(\theta_1)), \ldots,
\Phi^{-1}(G_{t,24}(\theta_{24}))),
\end{align}
where $\Phi$ is the CDF of the univariate standard normal distribution, and $\mathbf{\Phi}_\Sigma$ is the multivariate CDF of a Gaussian random vector with mean zero and correlation matrix $\Sigma$. The latter matrix has $24 \times 23/2 = 276$ unique elements, so that the Gaussian copula allows for considerable flexibility in modeling dependence, at the cost of parameter estimation uncertainty. In order to estimate the Gaussian copula, we use the rank correlation estimator (see, e.g., \citealt{Genest2007}).\\


In contrast to parametric copulas, non-parametric copulas do not assume a specific functional form for $C$. The use of nonparametric copulas in the context of multivariate forecast preprocessing was introduced by \cite{TheSchaakeShuffle}, who named the reordering idea after Dr. J. Schaake,  member of the National Weather Service Office of Hydrologic Development. \cite{schefzik2013} established the connection to empirical copulas. Our exposition of non-parametric copulas loosely follows \citet[Section 3.3]{schefzik2013}. Consider a training sample covering $m$ days of past point forecast errors for every hour $h$, i.e., 
\begin{align}
\{(\varepsilon_{t,1}, \ldots, \varepsilon_{t,24}): t = 1, \ldots, m\}.
\end{align}
We standardize these forecast errors by considering the corresponding quantile levels $\hat u_{t,h} = \hat G_{t,h}(\varepsilon_{t,h}) = \hat F_h(\hat z_t) \in [0,1]$\footnote{In the case of parametric (standard normal) margins, we have $    \hat{u}_{t,h}=\Phi(\hat{z}_{t,h}),$ where $\Phi$ is the CDF of the standard normal distribution. In the case of non-parametric margins, we have $    \hat{u}_{t,h}=\hat{F}_h(\hat{z}_{t,h})=\frac{\text{rk}(\hat{z}_{t,h})}{m+1},$ where $\text{rk}(\hat{z}_{t,h})$ denotes the rank of $\hat{z}_{t,h}$ within the standardized residuals for the $m$ days in the Dependence Learning Phase.}. Assume for simplicity that there are no ties, i.e., the training sample contains $m \times 24$ unique values of $\hat u_{t,h}.$ We denote the rank of $\hat u_{t,h}$ within $\hat u_{1,h}, \ldots, \hat u_{m,h}$ by $\text{rk}(\hat u_{t,h})$. That is, for the day $\underline{t}$ at which the smallest quantile level is observed, we have $\text{rk}(\hat u_{\underline{t},h}) = 1.$ Similarly, the day $\overline{t}$ with the largest quantile level yields 
$\text{rk}(\hat u_{\overline{t},h}) = m.$ The empirical copula based on a training sample of length $m$ is then given by
\begin{align}
\label{rankcop}
E_T(i_1/m, \ldots, i_{24}/m) = \frac{1}{m} \sum_{t=1}^m \mathbf{1}\left\{\text{rk}(\hat u_{t,1}) \le i_1, \ldots, \text{rk}(\hat u_{t,24}) \le i_{24}\right\},
\end{align}
for integers $0 \le i_1, \ldots, i_{24} \le m.$ For example, suppose that $i_1 = \ldots = i_{24} = 10.$ In this case, the above function yields the empirical frequency of the event that all $24$ quantile levels are simultaneously smaller or equal than their tenth smallest observed value. Clearly, setting $i_{1} = \ldots = i_{24} = 0$ yields a function value of zero, and setting $i_{1} = \ldots = i_{24} = m$ yields a function value of one.



\subsection{Forecasting Phase}
We next describe how to draw multivariate forecasts for the next day $t$. From the previous two steps, we have obtained estimates of the univariate distributions $G_{t,h}(x)$ and the copula of standardized residuals $\hat z_{t,h}$. We now combine these two pieces to construct the desired multivariate distribution.\\

Specifically, we first construct a probabilistic \textit{univariate} forecast distribution of size $m$, for each hour $h$ in the next day, given by 
        \begin{align}
        \reallywidehat{y_{t,h}}^{i}&= \underbrace{(\hat{y}_{t,h}+\hat{\mu}_{t,h})}_{\text{bias-corrected point forecast}} + \underbrace{\hat{F}_h^{-1}(i/(m+1))\times\hat{\sigma}_{t,h}}_{\text{predicted quantiles of forecast error}} 
        \label{rück}
    \end{align} for $i=1,\dots,m$, where 
$\hat{F}_h^{-1}$ is the inverse cumulative distribution function of the standardized residuals. In the case of parametric (Gaussian) marginal distributions, we have that $\hat{F}_h^{-1} = \Phi^{-1}$ is the quantile function of the standard normal CDF. Thus, Equation (\ref{rück}) generates a stylized sample by computing $m$ equally spaced quantiles of the forecast distribution. 
In the `raw-error' case without a time series model, we simply set $\hat{\mu}_{t,h}\equiv 0$ and $\hat{\sigma}_{t,h}\equiv 1$ in the formula above. \\

Up to now, the 24 ensembles represent the 24 marginal forecast densities for the hours, but lack the correct dependence structure. Therefore, the resulting $m$ univariate ensembles per hour have to be paired according to the learned copula to encode the desired dependence structure and a discrete representation of the copula is needed. Such a representation is given by a suitable rank matrix $\hat{R}$, containing the pairing of univariate ensembles. For the parametric copula approach, we derive the rank matrix from a random sample of size $m$ from the fitted parametric copula model. I.e., we assign ranks to $m$ random samples from the learned copula in each dimension starting with one for the smallest value. For the non-parametric approach, the rank-matrix $\hat{R}$ is calculated from the points of the empirical copula in the Dependence Learning Phase (see Equation \ref{rankcop}). We then sort the univariate ensemble forecasts for every hour according to the rank matrix $\hat{R}$ to adopt the dependence structure given by the copula. We illustrate this approach in Example \ref{examplecont} which complements the meteorological example of \citet[Section 3b]{TheSchaakeShuffle}.\\ 

\begin{example}
\label{examplecont}
To illustrate the method (and in particular, the Forecasting Phase), we next consider a toy example based on a Dependence Learning Phase of length $m = 7$, and four (instead of $24$) hours of the day. First, Table \ref{tab:univ_fc} illustrates the four univariate forecast distributions. Equation (\ref{rück}) together with $m = 7$ implies that we represent each distribution by a set of quantiles at levels $(1/8, \ldots, 7/8)$. The differences in the four univariate distributions (e.g., the lower median price forecast at 00:00 compared to 12:00) reflect intra-daily seasonality in energy prices. This seasonality is typically already reflected by the point forecasting model that we consider as an input to our method.\\
  \begin{table}[ht]
\centering
\begin{tabular}{rccccccc}
\toprule 
Quantile & && &&&\\
level & 12.5\% & 25\% & 37.5\%&50\%&62.5\% & 75\%& 87.5\% \\ \midrule
00:00 &  6.1 & 16.1 & 23.6 & 30.3   & 37.0  & 44.5  & 54.5\\
06:00 & 21.7 & 31.6 & 39.0 & 45.7   & 52.3  & 59.7  & 69.6\\
12:00 & 27.2 & 37.0 & 44.4 & 50.9   & 57.5  & 64.8  & 74.6\\
18:00 & 26.7 & 36.5 & 43.9 & 50.5   & 57.0  & 64.4  & 74.2\\

\bottomrule
\end{tabular}
\caption{Univariate forecast distributions in the toy example, separately for four hours $h$ (rows).}
\label{tab:univ_fc}
\end{table}

Table \ref{figrank22} illustrates an estimated rank matrix $\hat R$. For example, the table's first row shows the ranks for the first day of the Dependence Learning Phase. Prices were fairly low on that day: Among others, the price at 00:00 was the lowest recorded in the training sample, as compared to the prices at 00:00 for the other $m-1 = 6$ days in the training sample. The table also shows strong positive dependence of the prices across hours, as reflected by positive correlation of the ranks.\\

\begin{table}
\centering
\begin{tabular}{ccccc}
\midrule 
& 00:00 & 06:00 & 12:00 & 18:00 \\ \midrule
$t = 1$ & 1 & 2 & 1 & 2\\
$t = 2$ & 4 & 3 & 3 & 5\\
$t = 3$ & 5 & 4 & 7 & 7\\
$t = 4$ & 2 & 1 & 2 & 1\\
$t = 5$ & 3 & 5 & 5 & 6\\
$t = 6$ & 7 & 7 & 6 & 4\\
$t = 7$ & 6 & 6 & 4 & 3\\
\bottomrule
\end{tabular}
\caption{Rank matrix $\hat R$ in the toy example.}
\label{figrank22}
\end{table}
Finally, Table \ref{examplecont1} shows the multivariate forecast distribution that results from combining the univariate distributions in Table \ref{tab:univ_fc} with the rank structure from Table \ref{figrank22}. For example, the first draw of the multivariate forecast distribution consists of the lowest quantile for the price at 00:00 (given by $6.1$), the second-lowest quantile for the price at 06:00 (given by $31.6$), the lowest quantile for the price at 12:00 (given by $27.2$), and the second-lowest quantile for the price at 18:00 (given by $36.5$).

\begin{table}[ht]
    \centering
    \begin{tabular}{ccccc}\midrule 
Forecast & & & & \\
Draw \# & 00:00 & 06:00 & 12:00 & 18:00 \\ \midrule
1 & 6.1 & 31.6 & 27.2 & 36.5\\
2 & 30.3 & 39.0 & 44.4 & 57.0\\
3 & 37.0 & 45.7 & 74.6 & 74.2\\
4 & 16.1 & 21.7 & 37.0 & 26.7\\
5 & 23.6 & 52.3 & 57.5 & 64.4\\
6 & 54.5 & 69.6 & 64.8 & 50.5\\
7 & 44.5 & 59.7 & 50.9 & 43.9\\ \bottomrule
\end{tabular}
\caption{Final multivariate forecast distribution in the toy example.}
\label{examplecont1} 
\end{table}
\FloatBarrier

Observe that the number of elements is the same in Tables \ref{tab:univ_fc}, \ref{figrank22} and \ref{examplecont1}. At the same time, the interpretation differs across the three tables: While Table \ref{tab:univ_fc} represents quantile levels per time point in each row, rows in Table \ref{figrank22} represent ranks, and forecast draws in Table \ref{examplecont1}. This observation mirrors the simple yet clever construction behind the Schaake shuffle: Based on a dependence training sample of a given size (here, $m = 7$), it constructs a multivariate forecast sample of the same size. As a consequence, larger training samples yield larger and potentially more informative forecast distributions. On the other hand, short samples are quicker to adapt to possible structural breaks in dependence patters. In our empirical analysis below, we consider $m = 90$. 

\end{example}

\section{Results}
\label{sec:results}
\subsection{Forecast Evaluation}
For the evaluation of our forecast distributions, we follow the principle of \cite{PForecast1}, according to which a probabilistic forecast should maximize sharpness subject to calibration. Calibration means that, ideally, each observation should resemble a random draw from the predictive distribution. On the other hand, sharpness requires that the predictive distribution be as concentrated as possible. \textit{Proper scoring rules} \citep{PForecast2,PForecast3} assess sharpness and calibration simultaneously by assigning numerical predictive performance measures. A scoring rule is a function $S(F,y) \rightarrow \mathbb{R}$, where $F$ is the forecast distribution and $y$ the observed outcome. We consider scoring rules in negative orientation, i.e., a smaller score value indicates a better forecast. A scoring rule is called {proper} if stating the true forecast distribution yields the best expected score. We consider the \textit{Continuous Ranked Probability Score ({CRPS})} to evaluate the marginal forecast distributions. Specifically, we evaluate the forecast distribution for each of the $24$ horizons, and then compute the average score. Furthermore, we use the \textit{Energy Score ({ES})} to evaluate the multivariate forecast distribution.   
 We further apply the \cite{10.2307/1392185} test to assess the statistical significance of differences in forecast performance. Finally, we check the calibration of multivariate forecasts graphically with \textit{Average Rank Histograms} \citep{multrank}. We provide details on these evaluation methods in \ref{App:A}.

\subsection{Electricity Price Forecasting}
Our probabilistic forecasts are based on the rolling window scheme explained in Section \ref{rollwin}. We examine several variants of the proposed method to investigate the impact of different modeling choices on forecast performance. Additionally, we compare all settings to their simplified (independence) counterparts ignoring the dependence structure.

\subsubsection{Settings}
\label{sec:settings}

The variants we consider differ in their specification of the margin (standardizing forecast errors via a time series model versus using raw forecast errors; using a non-parametric versus Gaussian marginal distribution) and the copula (non-parametric versus Gaussian). For ease of presentation, we do not consider all possible configurations but focus on the settings listed in Table \ref{Tab:models}: The \textit{Schaake-NP} variant uses non-parametric margins and a non-parametric copula, as well as an AR(1)-GARCH(1,1) time series model for error standardization. The \textit{Schaake-P} variant uses Gaussian margins, a Gaussian copula, and an AR(1)-GARCH(1,1) for error standardization. Last, the \textit{Schaake-Raw} variant is based on raw forecast errors, i.e., without time series standardization, but is otherwise identical to the \textit{Shaake-NP} variant. As a simple evaluation of copula modeling performance, we compare each variant with its independence counterpart that uses the same marginal distributions of forecast errors for each hour $h$, but assumes them to be independent across hours. The AR-GARCH calculation is done using the `rugarch' package \citep{rugarch} in R \citep{R}, while everything else is implemented in Python 3 \citep{Python}.


\begin{table}[ht]
\centering
\small
\begin{tabular}{@{\extracolsep{4pt}}lrrrr@{}}

\hline
Setting& AR-GARCH&Marginal distributions&Dependence modeling\\
\hline

\\[-0.95em]
Schaake-NP  & Yes   & Non-parametric & Empirical copula\\
Schaake-P   & Yes   & Parametric    & Gaussian copula\\
Schaake-Raw & No    & Non-parametric & Empirical copula\\
I-NP        & Yes   & Non-parametric & Independence\\
I-P         & Yes   & Parametric    & Independence\\
I-Raw       & No    & Non-parametric & Independence\\
\hline
\end{tabular}
\caption{Settings considered for electricity price forecasting.  }
\label{Tab:models}
\end{table}

\subsubsection{Assessment}
\begin{figure}[ht]
\centering

\begin{subfigure}{.75\textwidth}
\centering
\includegraphics[width=\linewidth]{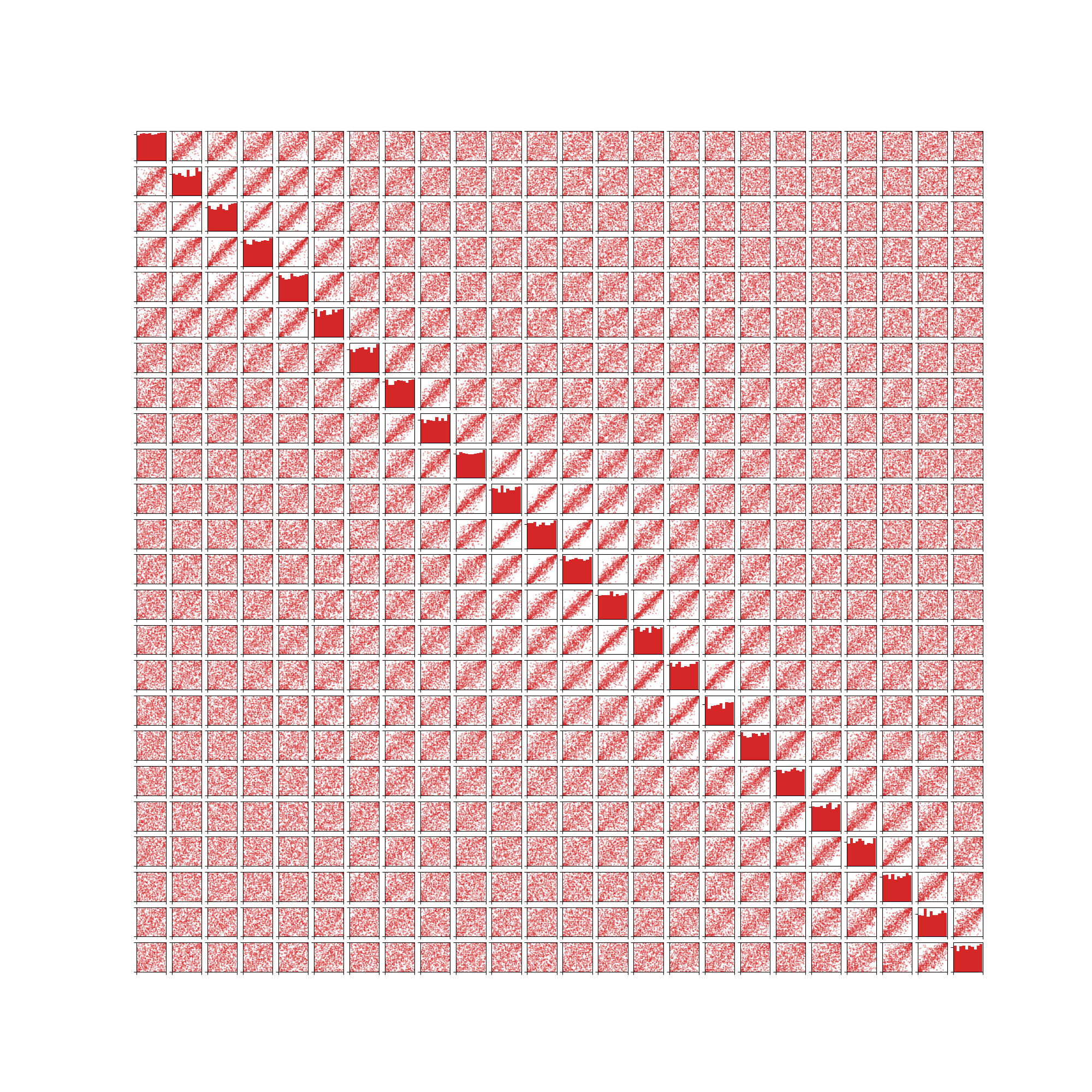}\vspace{-1cm}
\caption{Schaake-NP}
\end{subfigure}

\begin{subfigure}{.75\textwidth}
\centering
\includegraphics[width=\linewidth]{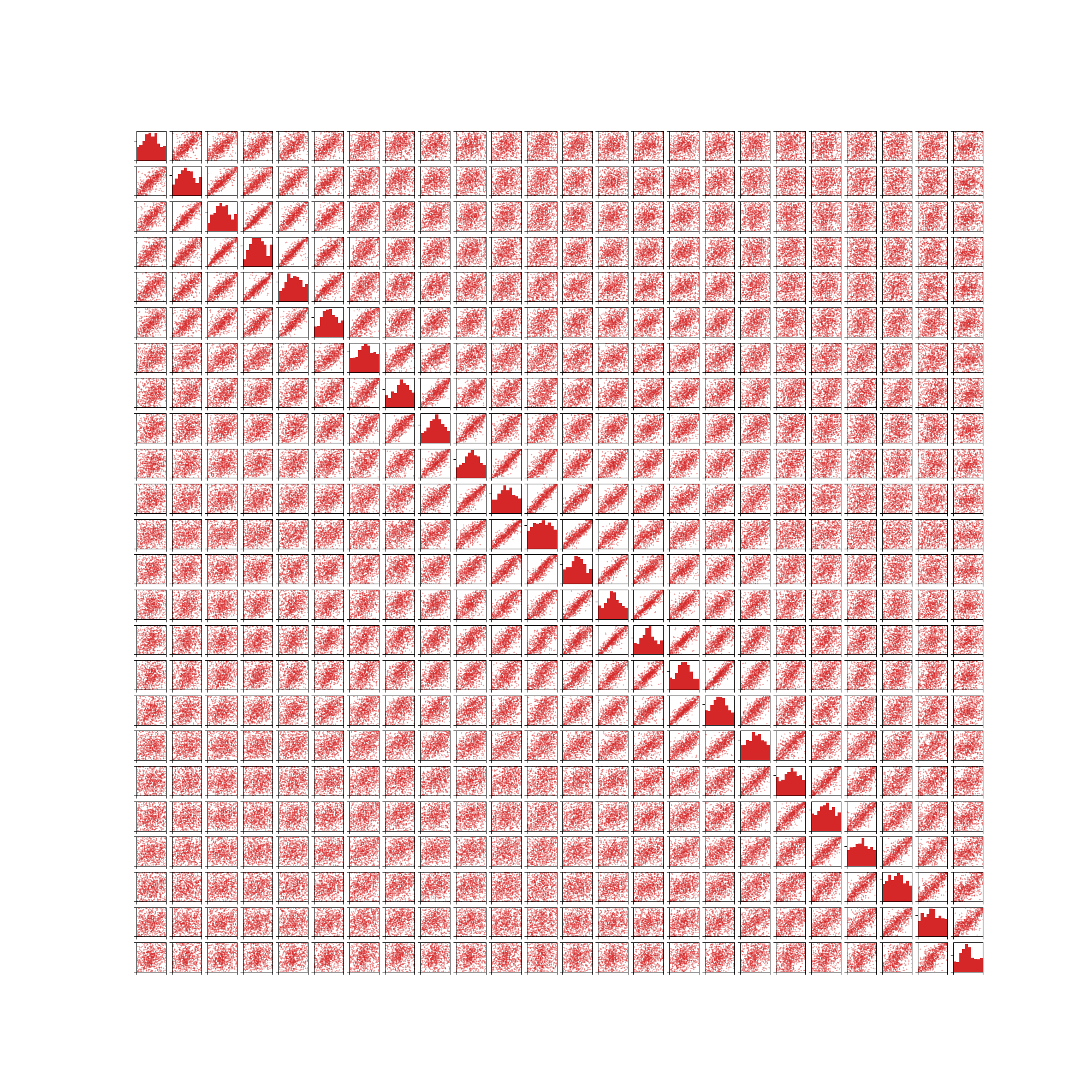}\vspace{-1cm}
\caption{Schaake-P}
\end{subfigure}




\caption{Verification rank histograms for the EPEX-DE data set. Diagonals: Univariate rank verification histograms. Off-diagonals: Bivariate scatterplots of realized quantiles.} \label{24h}
\end{figure}

Figures \ref{24h} (a), \ref{24h} (b) and \ref{24h} (c) provide illustrative calibration checks for the EPEX-DE data set. The univariate verification rank histograms for all 24 hours are displayed on the diagonal and the bivariate scatterplots of realized quantiles on the off-diagonals. Each point in the figure corresponds to one day in the forecasting phase. Figure \ref{24h} (a) displays the result for the Schaake-NP, while panel (b) displays the case of its parametric counterpart Schaake-P, and panel (c) refers to the case of raw errors (Schaake-Raw). The off-diagonal elements in Figures \ref{24h} (a) to (c) indicate a decreasing dependence for hours that are further apart, which seems plausible. The verification rank histograms (diagonal elements) suggest good calibration of the univariate forecast distribution for the non-parametric case (Figure \ref{24h} a), as indicated by uniform histograms. Univariate calibration seems to be slightly worse in the parametric case (diagonal elements of Figure \ref{24h} (b), where most of the histograms are hump-shaped. The margins of the non-parametric raw-error approach in Figure \ref{24h} (c) also seem to be well calibrated. Thus, the results suggest better calibration for the two non-parametric versions. Furthermore, satisfactory calibration of the raw error approach suggests that error post-processing via a time series model is not of much importance in this case study.\\ 

\addtocounter{figure}{-1}

\begin{figure}
    \centering
\begin{subfigure}{.75\textwidth}
\centering
\includegraphics[width=1\linewidth]{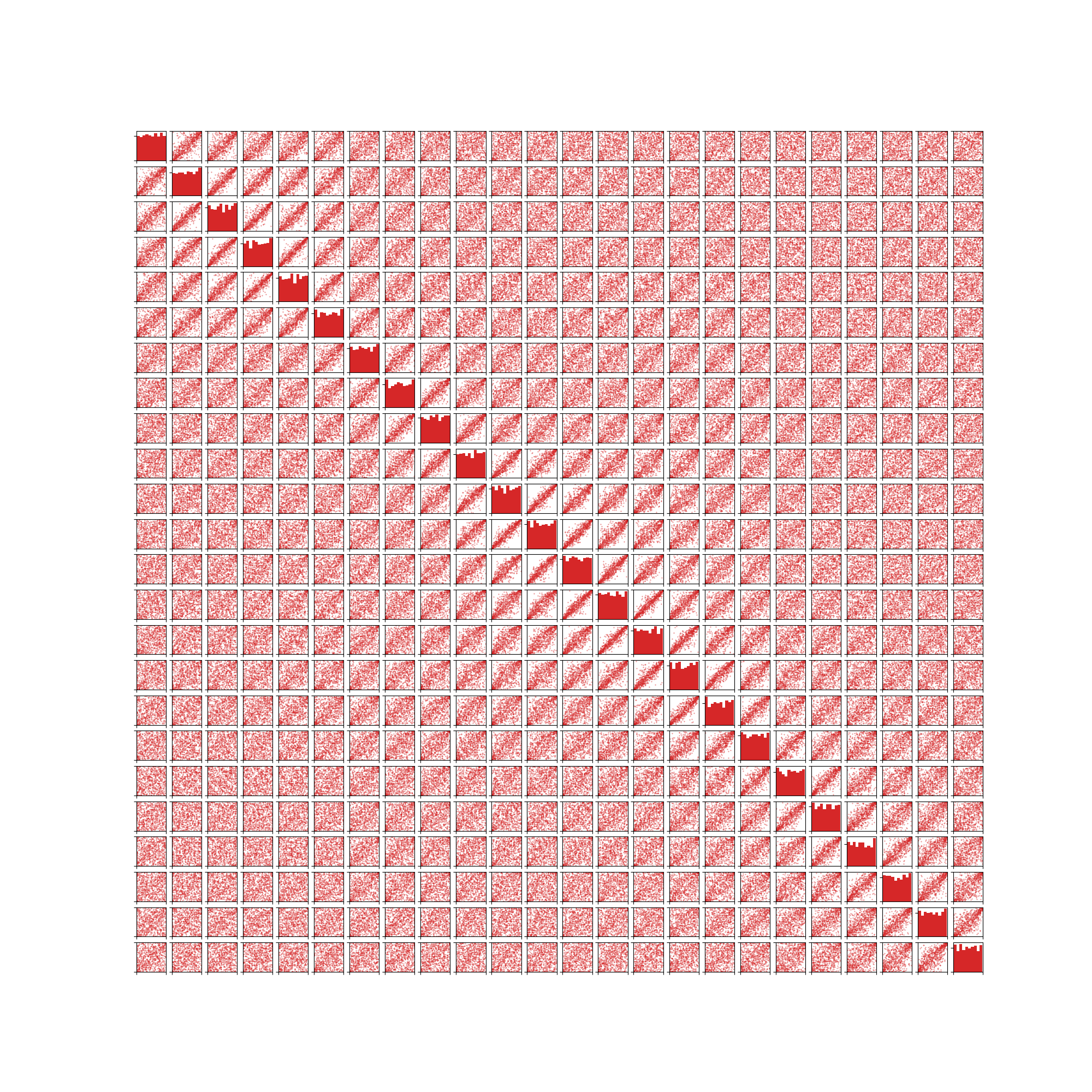}\\\vspace{-1cm}
{\scriptsize (c) Schaake-Raw}
\end{subfigure}\\[.4cm]
{\footnotesize Figure \ref{24h}, continued.}
\end{figure}

\begin{figure}[ht]
\centering

\begin{subfigure}{.33\textwidth}
\centering
\includegraphics[width=1\linewidth]{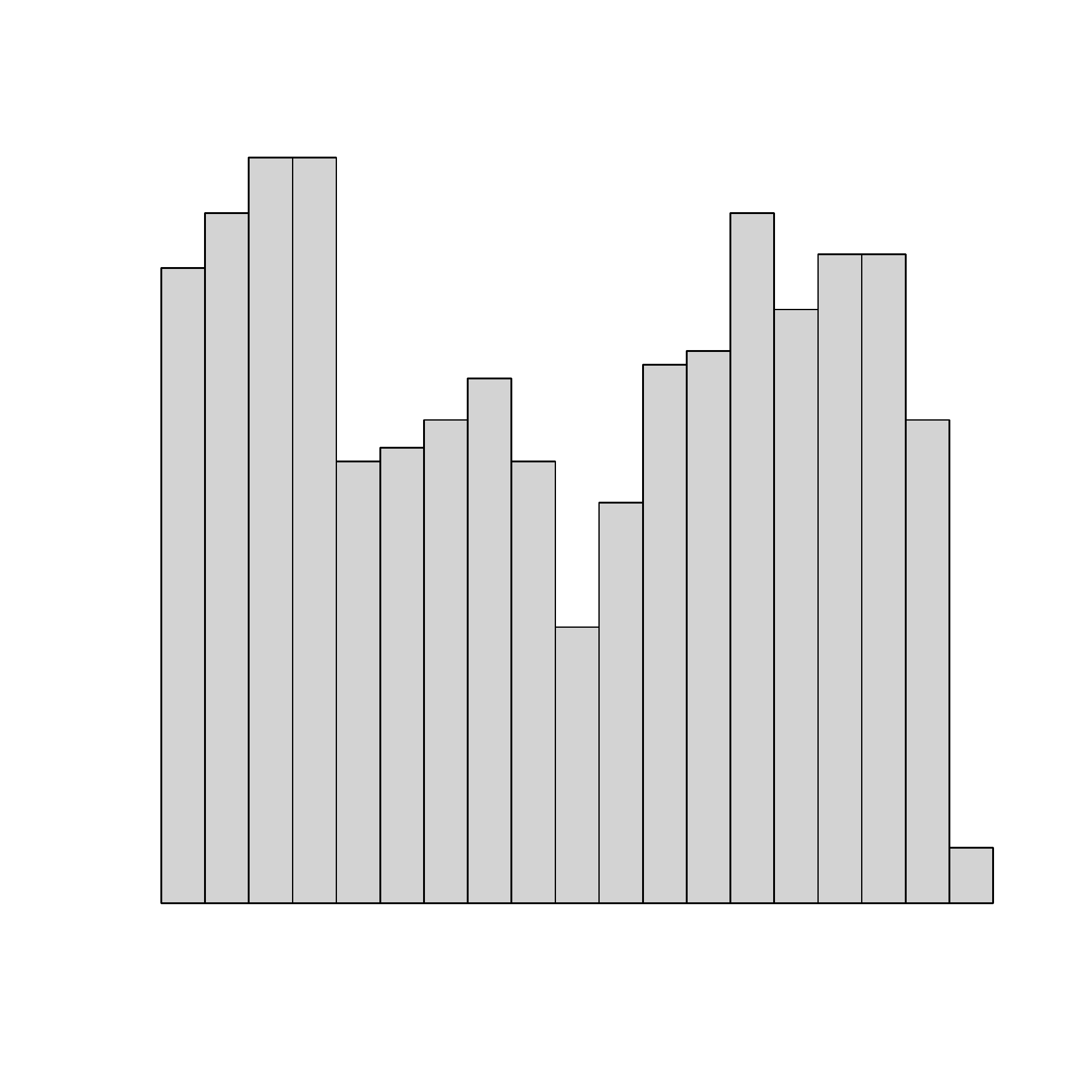}\vspace{-.4cm}
\caption{Schaake-NP}
\end{subfigure}%
\hfill
\begin{subfigure}{.33\textwidth}
\centering
 \includegraphics[width=1\linewidth]{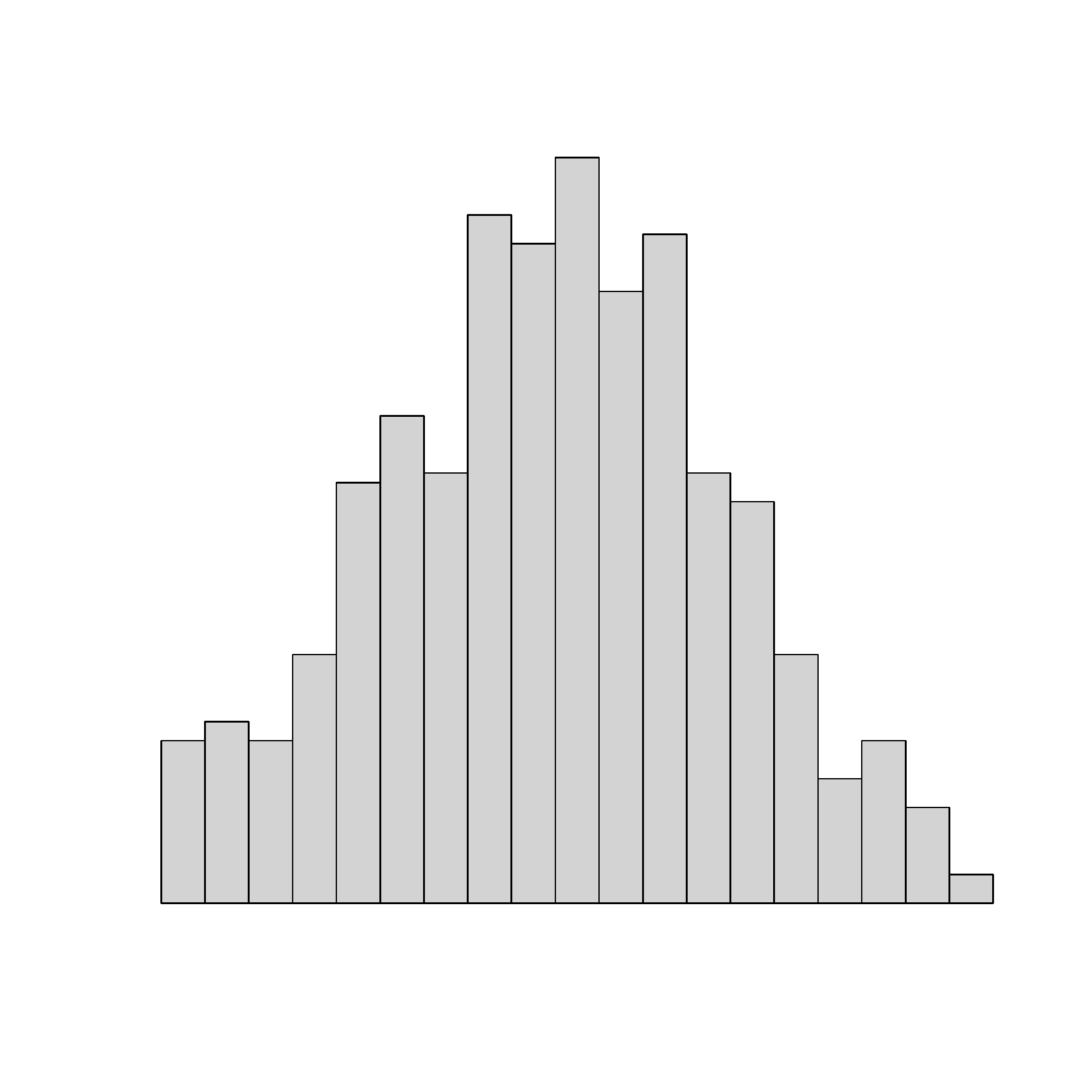}\vspace{-.4cm}
\caption{Schaake-P}
\end{subfigure}
\hfill
\begin{subfigure}{.33\textwidth}
\centering
\includegraphics[width=1\linewidth]{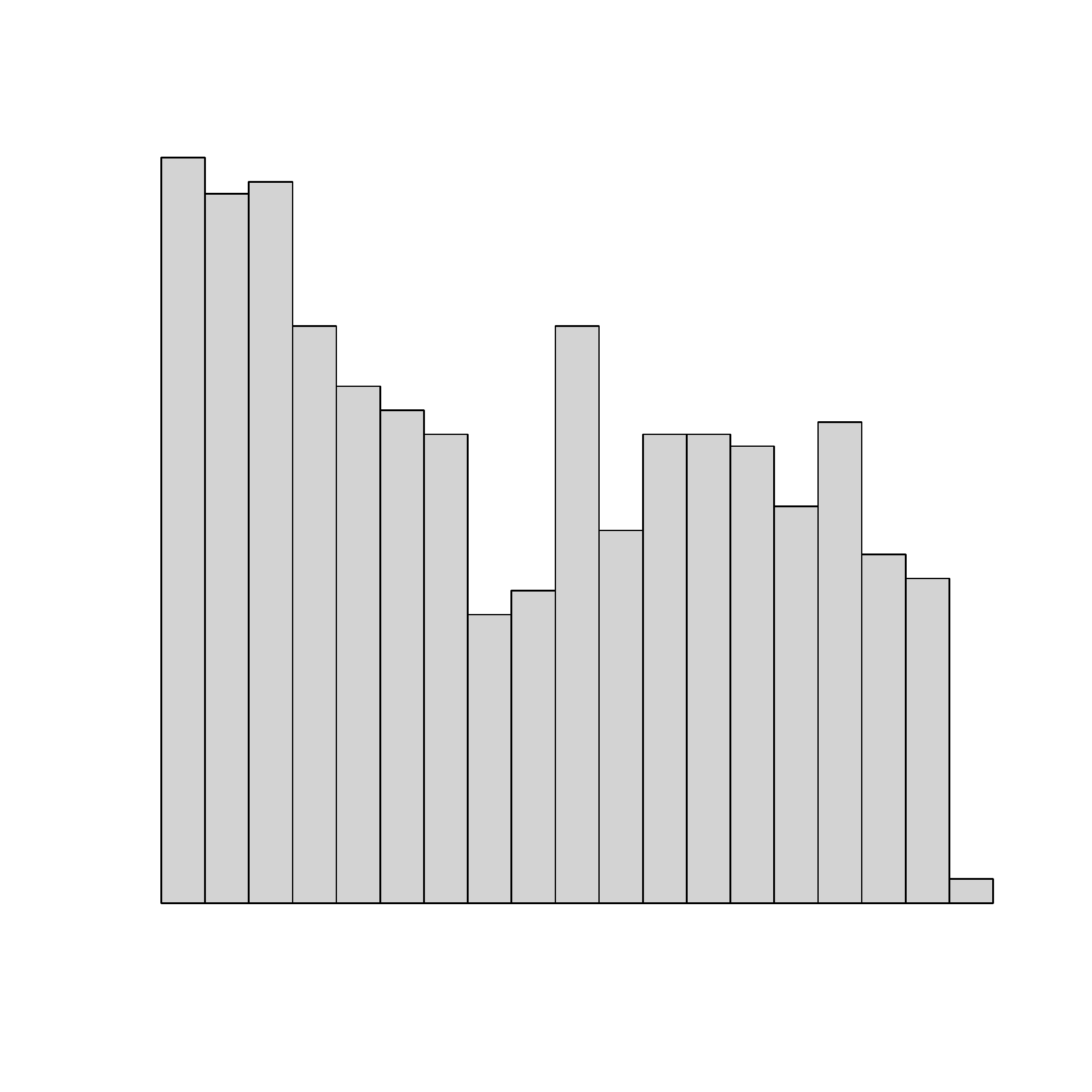}\vspace{-.4cm}
\caption{Schaake-Raw}
\end{subfigure}

\begin{subfigure}{.33\textwidth}
\centering
\includegraphics[width=1\linewidth]{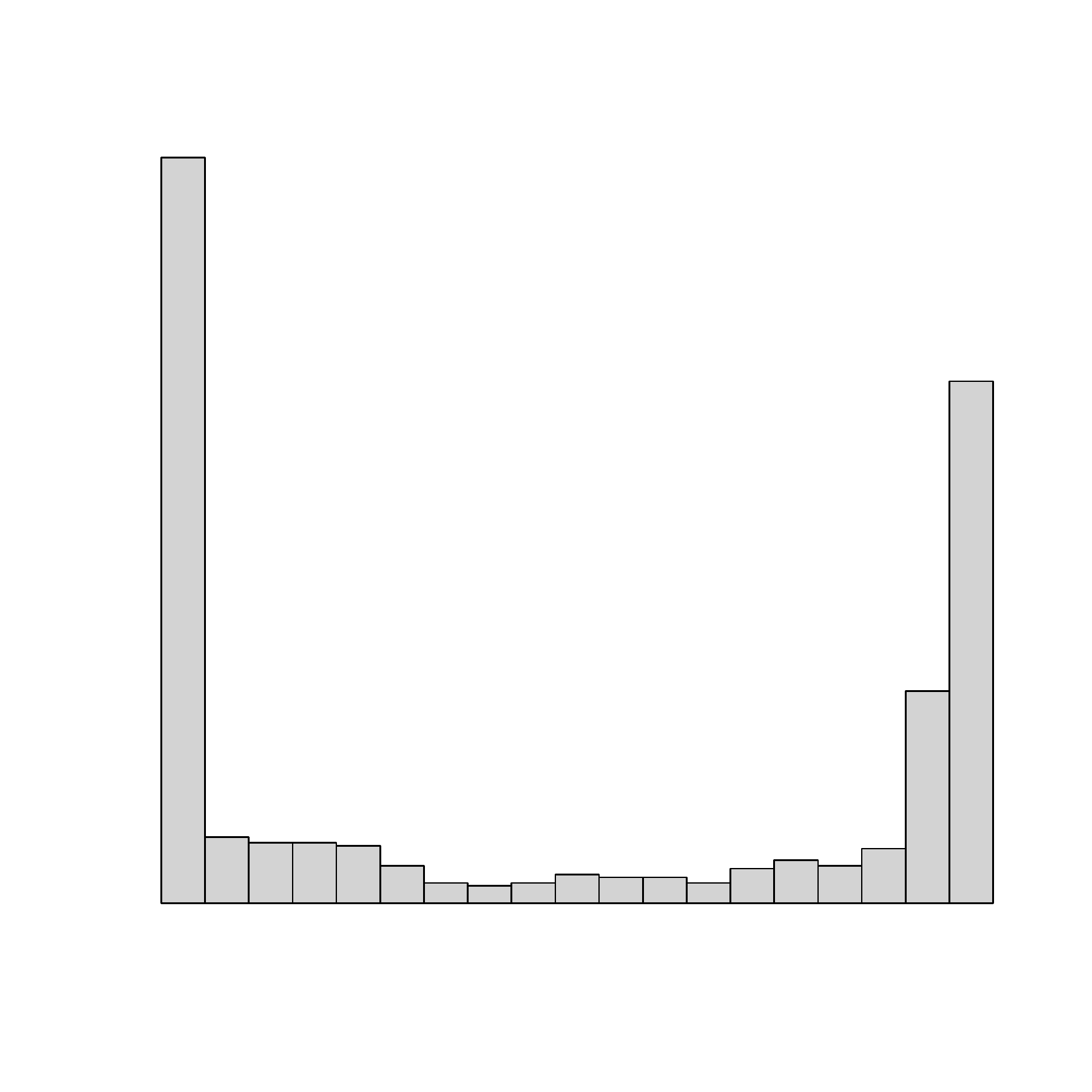}\vspace{-.4cm}
\caption{I-NP}
\end{subfigure}%
\hfill
\begin{subfigure}{.33\textwidth}
\centering
 \includegraphics[width=1\linewidth]{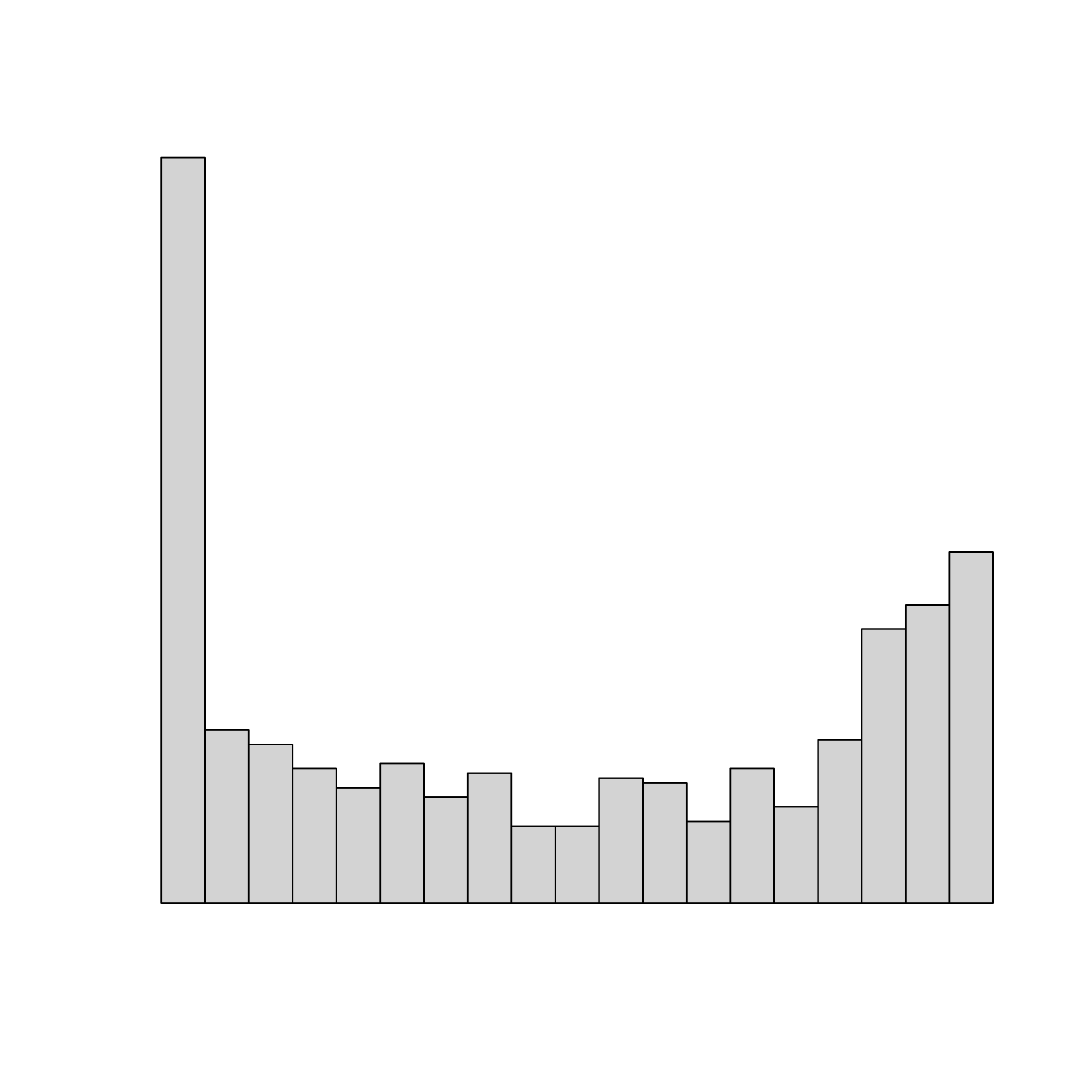}\vspace{-.4cm}
\caption{I-P}
\end{subfigure}
\hfill
\begin{subfigure}{.33\textwidth}
\centering
\includegraphics[width=1\linewidth]{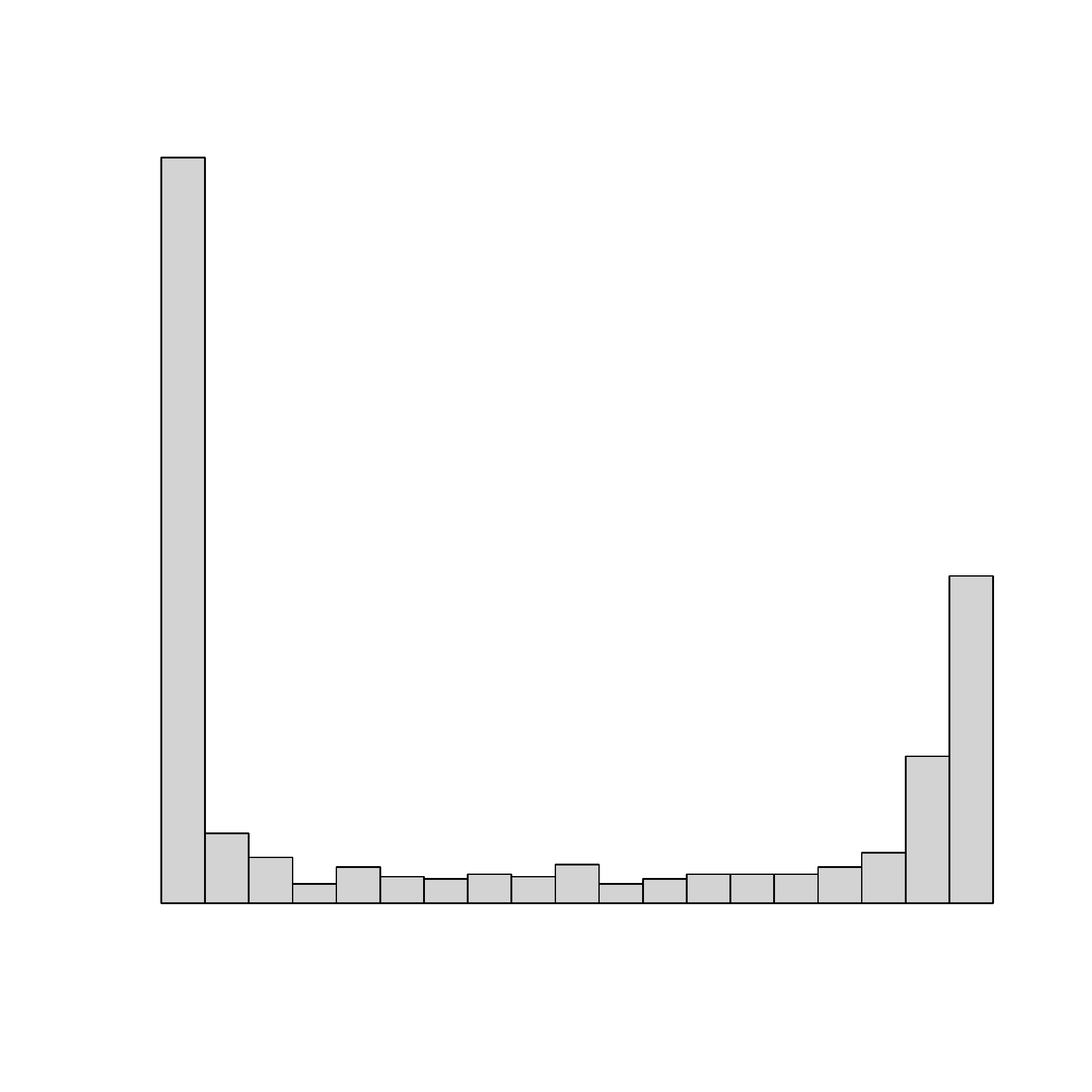}\vspace{-.4cm}
\caption{I-Raw}
\end{subfigure}

\caption{Average Rank Histograms for the DE dataset.}
\label{AV}
\end{figure}

Figure \ref{AV} shows the Average Rank Histograms for various forecast distributions. The nonparametric Schaake methods in panels (a) and (c) display slight signs of miscalibration (i.e., non-uniform histograms) which however seem rather unsystematic. The histogram for the parametric case in panel (b) is hump-shaped, which indicates under-dispersion and is in line with our observation in Figure \ref{24h} (b). The three na\"{i}ve independence counterparts in panels (d)-(f) clearly lack calibration, as indicated by distinct U-shaped patterns of the Average Rank Histograms.\\ 


\FloatBarrier

Finally, Table \ref{tabres} summarises the {Energy Score} and {CRPS} for all settings and the corresponding baselines in the five markets. We naturally observe the same values for the copula based approaches and their independence counterparts for the CRPS as they share the same marginal distributions. In terms of the Energy Score, the three copula based approaches consistently outperform their independence counterparts. Furthermore, we note a better performance for both non-parametric versions (Schaake-NP/Schaake-Raw) than for the parametric error distributions (Schaake-P). This result is in line with the lack of calibration of the Schaake-P method documented earlier.
Remarkably, the performance of the raw error approach (Schaake-Raw) is not systematically worse than that of the time series based non-parametric variant (Schaake-NP). For the DE, PJM and FR data sets, both variants perform similarly, and differences in score performance (Diebold-Mariano tests) are typically insignificant at the 5\% level. For the BE data set, which is challenging due to several positive spikes in prices \citep[Section 3.4]{lago2020forecasting}, Schaake-Raw outperforms Schaake-NP. This may be due to increased robustness of Schaake-NP in an unstable data environment. On the contrary, in the Nord Pole data, Schaake-NP outperforms Schaake-Raw. Overall, the competitive performance of the raw error approach indicates that there is little need of systematic error postprocessing. This suggests that the forecast models provided by \cite{lago2020forecasting} yield a good fit to the data, leaving little unmodeled heterogeneity in the models' forecast errors. The need for error postprocessing may be more pronounced in a setup with less sophisticated forecasting models. We consider such a case in the context of load forecasting below. 



\begin{table}[ht]
\centering
\begin{tabular}{@{\extracolsep{4pt}}lrrrr@{}}

\hline
Setting& Energy Score& CRPS&\multicolumn{2}{c}{$p$-values}\\
&&&(a)&(b)\\
\cline{1-1} \cline{2-2} \cline{3-3} \cline{4-5} 

\\[-0.95em]
 EPEX-DE \\
\hline
Schaake-NP  &  19.837&  3.297& & \\
Schaake-P  &  21.138&  3.611& $<$0.001 &$<$0.001\\
Schaake-Raw  &  19.787& 3.266& 0.685 &0.133\\
I-NP  &  20.398& 3.297& $<$0.001& \\
I-P  &  21.452& 3.611& $<$0.001 &$<$0.001\\
I-Raw  &  20.322& 3.266& $<$0.001& 0.133\\
\hline
\\[-0.95em]
\multicolumn{4}{@{}l}{Pennsylvania-New Jersey-Maryland (PJM) }\\ \hline    
Schaake-NP  &  14.594&  2.394& & \\
Schaake-P  &  15.318&  2.581& $<$0.001 &$<$0.001\\
Schaake-Raw  &  14.791& 2.436& 0.162 &0.047\\
I-NP  &  15.133& 2.394& $<$0.001& \\
I-P  &  15.687& 2.581& $<$0.001 &$<$0.001\\
I-Raw  &  15.426& 2.436& $<$0.001& 0.047\\
\hline
\\[-0.95em]

EPEX-BE   \\
\hline  
Schaake-NP  &  35.13&  5.503& & \\
Schaake-P  &  38.64&  6.134& $<$0.001 &$<$0.001\\
Schaake-Raw  &  33.488& 5.311& 0.012 &0.001\\
I-NP  &  36.423& 5.503& $<$0.001& \\
I-P  &  38.868& 6.134& $<$0.001 &$<$0.001\\
I-Raw  &  34.642& 5.311& 0.444& 0.001\\

\hline

\\[-0.95em]
EPEX-FR \\
\hline 
Schaake-NP  &  22.553&  3.307& & \\
Schaake-P  &  23.578&  3.515& 0.001 &$<$0.001\\
Schaake-Raw  &  21.203& 3.243& 0.097 &0.119\\
I-NP  &  23.088& 3.307& $<$0.001& \\
I-P  &  23.935& 3.515& $<$0.001 &$<$0.001\\
I-Raw  &  21.633& 3.243& 0.259& 0.119\\
\hline

\\[-0.95em]
Nord Pole  \\
\hline
Schaake-NP  &  9.729&  1.571& & \\
Schaake-P  &  10.437&  1.72& $<$0.001 &$<$0.001\\
Schaake-Raw  &  9.904& 1.626& 0.007 &$<$0.001\\
I-NP  &  10.179& 1.571& $<$0.001& \\
I-P  &  10.657& 1.72& $<$0.001 &$<$0.001\\
I-Raw  &  10.189& 1.626& $<$0.001& $<$0.001\\

\hline

\end{tabular}
\caption{Energy Score and CRPS for several variants of proposed method and five benchmark data sets. (a) $p$-value of Diebold-Mariano tests for equal Energy Score, compared to Schaake-NP. (b) $p$-value of Diebold-Mariano tests for equal CRPS, compared to Schaake-NP. Blank cells indicate that scores are equal by construction, so that the test is undefined.}

\label{tabres}
\end{table}

\FloatBarrier

\subsection{Load Profile Price Prediction Intervals} 
\label{slp_sec}
We illustrate the importance of the correct dependence structure in forecasting with the example of the Standard Load Profile (SLP) G0 from the \cite{BDEW}. To do so, we compare daily price forecasts for the SLP G0 from the Schaake-NP and Schaake-Raw variants with their independence counterparts over the time period from 04/01/2016 to 31/12/2017. We average the quarter-hourly SLP across seasons and aggregate it to full hours. The true price for the SLP of each day is then calculated with realized prices from the EPEX-DE dataset. Finally, we evaluate how often this price realizes within the 93.33\% quantile of the respective forecast distribution.\\
\begin{figure}[ht]
    \centering
\includegraphics[width=\linewidth]{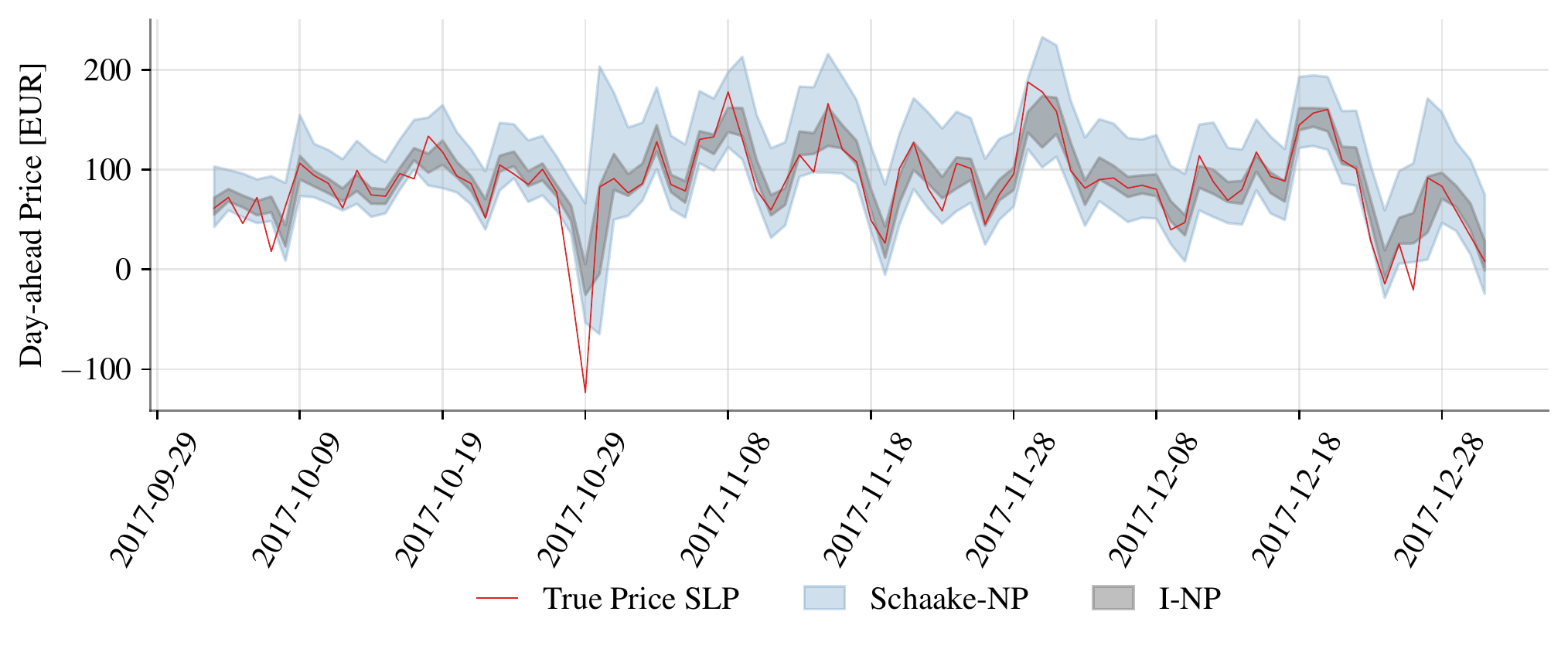}
    \caption{Prediction intervals and realized prices for SLP G0 from 3 October 2017 to 31 December 2017 (last 90 days of 2017).}
    \label{90Tage}
\end{figure}
The positive dependence between consecutive hours leads to wider prediction intervals for the SLP. Figure \ref{90Tage} illustrates this aspect for a three-month subsample of the data (for better visibility), whereas Figure \ref{SLPPriceinterval} in the Appendix presents the entire sample period. 
Table \ref{outside} summarizes the coverage rates of all methods over the two-year sample period. It shows that 93.27\% of the realized prices are within the projected 93.33\% prediction interval of the Schaake-NP model. For the Schaake-Raw model, the corresponding share is 91.76\%. Hence both models achieve coverage rates that are very close to nominal coverage, indicating good calibration. By contrast, the corresponding coverage rates of 54.81\% and 50.55\% for the independence assumption show that economic uncertainty cannot be well approximated without considering the dependence structure. 
\begin{table}[ht]
    \centering
    \begin{tabular}{ccc}
    \hline
         &Independence Assumption& Copula Model (Schaake)\\
        \hline
         AR(1)-GARCH(1,1)      &54.81\%    & 93.27\%           \\
        Raw     &50.55\%    & 91.76\%            \\
        \hline
    \end{tabular}
    \caption{Coverage of prediction intervals (nominal level: 93.33\%) for realized SLP prices.}
    \label{outside}
\end{table}
As a simple alternative method, one could compute a prediction interval from past price realizations. In the present case, using this method with a sample of $90$ past prices (in line with the choice for the Dependence Learning Phase) also yields a satisfactory coverage rate of 88.95\%. Note, however, that the simple method is specific to the particular load profile considered here. By contrast, the Schaake methods can be adapted to any given load profile. 

\FloatBarrier

\subsection{Forecasting intraday load data}
We finally give a short example for other operational scenarios where our proposed method can be applied. Instead of hourly prices, this example uses intra-day load data of the entso-e transparency platform for Germany from 2016 to 2020 (\url{https://transparency.entsoe.eu/dashboard/show}). We average the day-ahead quarter-hourly load forecasts ([6.1.B]) and actual loads ([6.1.A]) over each hour, hence resulting in $24$ hourly forecast error observations, in line with the setup of the price data considered earlier. We choose this data set to substantiate the use of time series models as an optional building block within the proposed method. In contrast to the point forecast errors for the price data above, point forecast errors for the load data contain time series structure with a marked seasonal component (see, e.g., the analysis in \cite{MACIEJOWSKA2021105273}). Without claiming to be optimal, we tackle this stylized fact by using a SARIMA(1,0,0)(1,0,0,7) time series with one AR parameter and one seasonal AR parameter with a lag of one week, i.e., 7 days for each hourly error time series (Schaake-NP). See \citealt[Section 9.9]{HA21} for an introduction to SARIMA models. We display the results for the Energy Score and CRPS for the years 2019 and 2020 in Table \ref{entso_e table}. 

\begin{table}[ht]
    \centering
    \begin{tabular}{lccccc}
\hline
Setting& Energy Score& CRPS&\multicolumn{2}{c}{$p$-values}\\
&&&(a)&(b)\\
\cline{1-1} \cline{2-2} \cline{3-3} \cline{4-5} 
 \hline
Shaake-NP  &  5854.619&  & 1047.883&\\
Shaake-Raw  &  13412.822&  & 2614.415& $<0.001$& $<0.001$\\
I-NP  &  6082.442&  & 1047.883& $<0.001$ &\\
I-Raw  &  13734.781&  & 2614.415& $<0.001$ & $<0.001$\\
\hline
        

    \end{tabular}
    \caption{Energy-Score and CRPS for forecasting German load data from 01 January 2019 to 30 December 2020. (a) $p$-value of Diebold-Mariano tests for equal Energy Score, compared to Schaake-NP. (b) $p$-value of Diebold-Mariano tests for equal CRPS, compared to Schaake-NP. Blank cells indicate that scores are equal by construction, so that the test is undefined.}
    \label{entso_e table}
\end{table}
\FloatBarrier

The table indicates that raw error approaches are clearly inferior to SARIMA based error postprocessing. Thus, in contrast to the case of price data considered abover, there appears to be a clear need for error postprocessing in this example. We conjecture that this result is driven by the lower sophistication of the (unknown) load forecasting model, as compared to the forecasting models provided by \cite{lago2020forecasting} that we considered for the energy price data. 


\section{Conclusion}
\label{sec:conclusion}
This paper proposes a postprocessing method to create multivariate forecast distributions for day-ahead electricity prices out of any point forecasting model. The method is motivated by two main aspects: First, many sophisticated point forecasting models have been introduced in the literature, so that we may take the availability of a `good' point forecasting model as given. Second, many economic decision require probabilistic multivariate forecasts which are hardly available at present.\\

Our method exploits several time series of univariate point forecast errors and creates a multivariate forecast distribution which inherits their dependence structure. In a case study on energy price forecasting based on benchmark models of \cite{lago2020forecasting}, a simple raw-error variant of the method performs well. This result indicates that the point forecast errors contain little exploitable structure, so that using a time series model for these errors is not necessary. In an additional case study on load forecasting, the point forecast errors contain pronounced seasonal patterns, and removing these patterns via a time series model leads to clear gains in forecasting performance. Throughout our empirical analysis, simple nonparametric techniques (for estimating the marginal distribution and the copula) outperform parametric ones based on normality assumptions.\\




\newpage

\appendix

\section{Algorithmic Description of Proposed Method}

Here we present algorithmic descriptions of the proposed method, in the version with a time series model (Algorithm 1) and in the raw-error version without a time series model (Algorithm 2). For ease of presentation, we omit time indices in this section; for example, we denote the price for hour $h$ of day $t$ by $y_h$ instead of $y_{t,h}$. 

\label{App:A}
\begin{algorithm}[ht]
 \footnotesize
  \algsetup{linenosize=\tiny}
  \scriptsize
\SetAlgoLined
\KwResult{Simulated multivariate forecast distribution for day ahead electricity prices}
\KwInput{History of point forecasts for day ahead electricity prices}
\For{each day in Error Learning Phase}{
 \For{each hour h}{
   Fit time series model to errors in learning data $\epsilon_{h}=y_{h} -\hat{y}_{h}$ \\
     Standardize realized forecast error by $\hat z_{h}=\frac{\epsilon_h -\hat{\mu}_h}{\hat{\sigma}_{h}}$ with $\hat{\mu}_h$ and $\hat{\sigma}_{h}$ obtained from time series model and save last 90 of the resulting standardized error $\hat{z}_h$ in $\hat{F}_h$
 }}
\For{each day in Dependence Learning Phase}{
 \For{each hour h}{
 \eIf{parametric margins==True}{
   Calculate realized quantile $\hat u_h$ of $\hat z_h$ in specified distribution, e.g., $N(0,1)$, by $  \hat u_h=\Phi(\hat{z}_{h}) $ and save in $C$  \
   }{
   Calculate realized quantile $\hat{u}_h$ of $\hat{z}_h$ in empirical distribution $\hat{F}_h$ by $\hat u_h=\hat{F}_h(z_h)$ and save in $C$ } }}
 \For{each day t in Forecasting Phase}{
 \For{each hour h}{
 Create $m$ univariate samples representing each $\frac{1}{m+1}^{th}$ quantile \\
  \eIf{parametric margins==True}{
     use specified distribution, e.g., $N(0,1)$ and its inverse CDF $\Phi^{-1}$:\\
  \For{each ensemble member i}{  $\reallywidehat{y_{h,ens}}^{(i)}=(\hat{y}_h+\hat{\mu}_h)+\Phi^{-1}(i/(m+1))\times\hat{\sigma}_h$}}
  {use empirical CDF $\hat{F}_h$ and its inverse CDF $\hat{F}_h^{-1}$:\\
   \For{each ensemble member i}{  $\reallywidehat{y_{h,ens}}^{(i)}=(\hat{y}_h+\hat{\mu}_h)+\hat{F}_h^{-1}(i/(m+1))\times\hat{\sigma}_h$}}
  
 }\eIf{parametric dependence==True}{
Fit specified parametric copula model to $C$\\
   Sample $m$ times from parametric copula model of $C$ and create rank-matrix\\
   }{Create rank matrix out of saved values from $C$\\
   
  }Pair up univariate forecast ensembles according to the rank-matrix of $C$.}
 \caption{Schaake shuffle with time series model (Schaake-NP/Schaake-P)}
 \label{algo1}
\end{algorithm}
\FloatBarrier
\newpage
\begin{algorithm}[ht]
 \footnotesize
  \algsetup{linenosize=\tiny}
  \scriptsize
\SetAlgoLined
\KwResult{Simulated multivariate forecast distribution for day ahead electricity prices}
\KwInput{History of point forecasts for day ahead electricity prices}
\For{each day in Error Learning Phase}{
 \For{each hour h}{
   Save last 90 errors in learning data $\epsilon_{h}=y_{h} -\hat{y}_{h}$ \\
 }}
\For{each day in Dependence Learning Phase}{
 \For{each hour h}{
   Calculate realized quantile $\hat{u}_h$ of $\epsilon_h$ in empirical distribution $\hat{F}_h$ by $\hat u_h=\hat{F}_h(\epsilon_h)$ and save in $C$  }}
 \For{each day t in Forecasting Phase}{
 \For{each hour h}{
   \For{each ensemble member i}{  $\reallywidehat{y_{h,ens}}^{(i)}=(\hat{y}_h+\hat{\mu}_h)+\hat{F}_h^{-1}(i/(m+1))\times\hat{\sigma}_h$}
 }Pair up univariate forecast ensembles according to the rank-matrix of $C$.}
 \caption{Schaake shuffle without time series model (Schaake-Raw)}
 \label{algo2}
\end{algorithm}
\FloatBarrier

\section{Details on Forecast Evaluation }
\label{eVALUATION}
\subsection{Assessment of Calibration}
We assess univariate calibration by examining \textit{verification rank histograms} \citep[see e.g.][]{MR2434318}, separately for each hour $h = 1, 2, \ldots, 24$. To describe the verification rank histogram for hour $h$, let $R_{t,h} = 1 + \sum_{i=1}^{90} \mathbf{1}(\hat y_{t,h}^i < y_{t,h})$ denote the rank of the realization $y_{t,h}$ in the merged sample $\{y_{t,h}, \{\hat y_{t,h}^i\}_{i=1}^{90}\}$ that contains the realization and the forecast sample of size $90$. If the forecast for hour $h$ is calibrated, then the distribution of $\{R_{t,h}\}_{t=1}^T$ is uniform, indicating that observed prices behave like random draws from the (hour-specific) forecast distribution. By contrast, a U-shaped distribution indicates that observed prices are often higher or lower than expected by the forecast distribution, implying that the latter is too narrow (or overconfident).\\

To assess multivariate calibration, we use the \textit{Average Rank Histogram} \citep{multrank} which has a relatively simple interpretation that is similar to the univariate case. The Average Rank Histogram is based on the average rank across hours, given by $\text{AvgRank}_t = (\sum_{h=1}^{24} R_{t,h})/24$. If the forecast is calibrated, the latter series is uniformly distributed.



\subsection{CRPS}
To assess the univariate forecasts, we use the \textit{Continuous Ranked Probability Score (CRPS)}, which we average over time and across all 24 hours $h$. For a univariate forecast distribution $F$ (represented by its cumulative distribution function, or CDF) and an outcome $y$, the CRPS is defined by
\begin{align}
\begin{split}
     CRPS(F,y)&=\int \bigl(F(x)-\mathbf{1}(x\geq y)\bigl)^2 dx\\
    &= \mathbb{E}_F|Y-y|-\frac{1}{2}\mathbb{E}_F|Y-Y'|,
\end{split}
\end{align} where $Y$ and $Y'$ are independent random variables distributed according to $F$, and $ \mathbf{1}(x\geq y)$ denotes the indicator function which equals one if $x \geq y$ and zero otherwise. We aggregate the $T \times 24$ score values (for each day and hour) by computing their mean. 

\subsection{Energy Score}
The multivariate extension of the CRPS, the \textit{Energy Score (ES)}. For a multivariate forecast distribution $F$, the ES is given by
\begin{align*}
    ES(F,y)=\mathbb{E}_F||Y-y||-\frac{1}{2}\mathbb{E}_F||Y-Y'||.
\end{align*}
 For an ensemble forecast $F_{ens}$ with members $\hat{y}^{(1)},\dots,\hat{y}^{(m)}$ it reduces to 
 \begin{align}
     ES(F_{ens},\mathbf{y})= \frac{1}{m}\sum_{k=1}^{m}||\hat{y}^{(k)}-y||-\frac{1}{2m^2}\sum_{l=1}^{m}\sum_{k=1}^{m}||\hat{y}^{(l)}-\hat{y}^{(k)}||,
 \end{align} see  \cite{MR2434318}. We average the ES across the $T$ days of the evaluation sample. 
 
 \subsection{Diebold-Mariano-Test}
We apply the \cite{10.2307/1392185} test for testing whether differences in forecast performance are statistically significant. The test considers the null hypothesis that $\mathbb{E}(S_{1,t}-S_{2,t}) = 0,$ where $S_{1t}$ and $S_{2t}$ denote the scores of forecasting methods $1$ and $2$ on day $t$. In our case, the scores correspond to either the CRPS or the Energy Score. The Diebold-Mariano test is a $t$-test with test statistic given by \begin{align*}
   \frac{\bar\delta_{12}}{\hat \sigma(\bar\delta_{12})}
\end{align*} 
where $\bar \delta_{12} = T^{-1} \sum_{t=1}^T (S_{1,t}-S_{2,t})$ is the mean score difference, and
$\hat\sigma(\bar\delta_{12})$ is its estimated standard deviation. The test statistic is standard normally distributed under the null hypothesis. Since we focus on one-day ahead forecasts, we set $\hat\sigma(\bar\delta_{12})$ equal to the sample standard deviation.

\newpage
\section{Prediction Intervals in Load Profile Example}

\begin{figure}[ht]
    \begin{centering}
\includegraphics[width=1\linewidth]{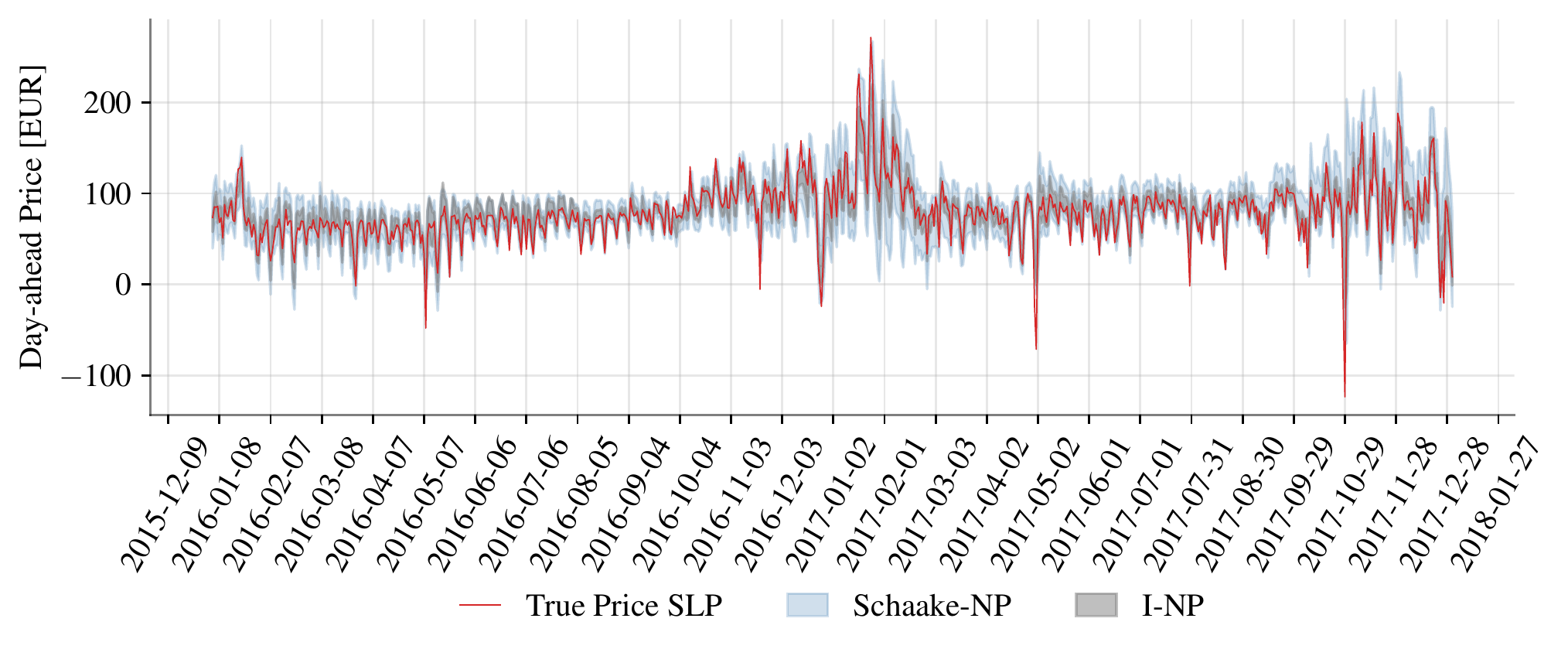}
    \caption{Prediction intervals and realized prices for SLP G0 from 4 January 2016 to 31 December 2017. }
    \label{SLPPriceinterval}
     \end{centering}
\end{figure}

Figure \ref{SLPPriceinterval} displays prediction intervals in the load profile example of Section \ref{slp_sec} for the full sample period. It complements Figure \ref{90Tage} in the main text that zooms in on the last 90 days of the sample period.

\FloatBarrier

\section{Code}

Computer code is available at the following link: \url{https://github.com/FabianKaechele/Energy-Schaake}

\section*{Funding}
This research did not receive any specific grant from funding agencies in the public, commercial, or not-for-profit sectors.
\section*{Declaration of Interest}
No conflict of interests to declare.
\FloatBarrier
\bibliographystyle{elsarticle-harv.bst}
\bibliography{manuscript}
\end{document}